\documentclass[12pt,a4paper]{article}
\usepackage{a4wide}
\usepackage{graphicx}
\usepackage[T1]{fontenc}
\usepackage[utf8]{inputenc}
\usepackage{textcomp}
\usepackage[sc,osf]{mathpazo}
\usepackage{latexsym,amsthm,amsfonts,amsmath,mathrsfs,amssymb}
\usepackage{dsfont}
\usepackage{accents}
\usepackage{ulem}
\usepackage[nosort]{cite}
\usepackage{booktabs} 
\usepackage[unicode,implicit]{hyperref}
\hypersetup{%
  pdftitle    = {Singular limits of spacetimes and their isometries}
  pdfkeywords = {~},
  pdfauthor   = {Eric Bergshoeff, Javier Matulich, Tom\'as Ort\'{\i}n},
  plainpages  = true,
  colorlinks  = true,
  citecolor   = blue,
  urlcolor    = red,
  linkcolor   = black
}
\newcommand{\hepth}[1]{{\tt
\href{http://www.arXiv.org/abs/hep-th/#1}{hep-th/#1}}}

\newcommand{\arxiv}[1]{{\tt arXiv:\href{http://www.arXiv.org/abs/#1}{#1}}}
\newcommand{\doi}[1]{{\tt DOI:\href{http://dx.doi.org/#1}{#1}}}

\allowdisplaybreaks

\makeatletter
\@addtoreset{equation}{section}
\makeatother

\begin{document}

\begin{flushright}
\small
IFT-UAM/CSIC-23-103\\
September 21\textsuperscript{st}, 2023\\
\normalsize
\end{flushright}

\vspace{1cm}

\begin{center}

  {\Large {\bf On the symmetries of singular  limits of spacetimes}}

\vspace{1.5cm}

\renewcommand{\thefootnote}{\alph{footnote}}

{\sl\large Eric Bergshoeff,}$^{1,}$\footnote{Email: {\tt
    e.a.bergshoeff[at]rug.nl}} {\sl\large Javier
  Matulich}$^{2,}$\footnote{Email: {\tt  javier.matulich[at]csic.es}}
{\sl\large and Tom\'as Ort\'{\i}n}$^{2,}$\footnote{Email: {\tt  tomas.ortin[at]csic.es}}

\setcounter{footnote}{0}
\renewcommand{\thefootnote}{\arabic{footnote}}
\vspace{1cm}

${}^{1}${\it Van Swinderen Institute, University of Groningen\\
Nijenborgh 4, 9747 AG Groningen, The Netherlands}\\

\vspace{.5cm}

${}^{2}${\it Instituto de F\'{\i}sica Te\'orica UAM/CSIC\\
C/ Nicol\'as Cabrera, 13--15,  C.U.~Cantoblanco, E-28049 Madrid, Spain}\\

\vspace{1cm}


{\bf Abstract}
\end{center}
\begin{quotation}
  {\small We consider spacetime metrics with a given (but quite generic)
    dependence on a dimensionful parameter such that in the $0$ and $\infty$
    limits of that parameter the metric becomes singular. We study the
    isometry groups of the original spacetime metrics and of the singular
    metrics that arise in the limits and the corresponding symmetries of the
    motion of $p$-branes evolving in them, showing how the Killing vectors and
    their Lie algebras can be found in general. We illustrate our general
    results with several examples which include limits of anti-de Sitter
    spacetime in which the holographic screen is one of the singular metrics
    and of $pp$-waves.}
\end{quotation}

\newpage
\pagestyle{plain}

\tableofcontents


\section{Introduction}

In the past few years, there has been a lot of activity in studying
non-relativistic, or, more generally, non-Lorentzian gravity and string
theories.\footnote{In this work we will generically call ``non-relativistic''
  or ``non-Lorentzian'' theories which do not exhibit the full invariance
  corresponding to their space and time dimension. Some of these theories may
  be invariant under a lower-dimensional Lorentz group and, therefore, may
  still be ``relativistic'' or ``Lorentzian'' in a restricted sense.} For some
recent reviews, see \cite{Oling:2022fft,Bergshoeff:2022eog,Hartong:2022lsy}.
The singular geometries underlying these gravity and string theories are quite
different from the regular geometries underlying the relativistic parent
theory in the sense that the regular geometry can be endowed with a single
metric whereas the singular geometry is characterized by two separate metrics.
A characteristic feature of taking the singular limit of a regular geometry
leading to a degenerate geometry with two separate metrics is that divergences
arise that need to be taken care of. These infinities are essential to make
the transition between the two types of geometries possible.

A noteworthy feature of coupling extended objects, such as particles, strings,
membranes etc., to a geometric background is that, whereas all extended
objects couple to the same Riemannian background in the relativistic case, in
the non-Lorentzian case there are different non-relativistic backgrounds to
which the extended objects of different worldvolume dimensions will naturally
couple.  Consider a $p$-dimensional extended object (a ``$p$-brane'') moving
in a $d$-dimensional background, with $p+1$ (``longitudinal'') spacetime
directions parallel to the object's worldvolume and $d-p-1$ directions
transverse to it. Then, a natural non-relativistic background to which such a
$p$-brane may couple is a degenerate geometry foliated by spatial
submanifolds of dimension $d-p-1$. The transverse and longitudinal directions
are inequivalent in the sense that, under a boost transformation, a transverse
direction transforms into a longitudinal direction but not the other way
round. To obtain these different foliated geometries from the same Riemannian
geometry one needs to define different so-called ``$p$-brane limits'' in which
the longitudinal and transverse directions are treated in a different way. For
a recent discussion of such $p$-brane limits, see \cite{tobesubmitted}.

In general, to define a limit, be it singular or not, we must start by
redefining the fields (and possibly other constants) of the relativistic
theory using a dimensionful ``contraction parameter''
$\rho$.\,\footnote{Strictly speaking the true contraction parameter $\lambda$
  should be dimensionless. Such a parameter can be obtained by redefining
  $\rho \to \lambda \rho$. With a slight misuse of language we will often call
  $\rho$ the contraction parameter.} This contraction parameter can be
anything: it could be the velocity of light $c$\footnote{This is the
  conventional choice in the simplest non-relativistic limits.} but it could
also be some radial parameter $R$ or the cosmological constant.  The
redefinition is done in such a way that no new fields are introduced and the
redefinition is invertible. The redefined theory is still relativistic, but we
will call the redefined fields the would-be non-Lorentzian fields in the sense
that they will become the fields of the non-Lorentzian theory {\it after}
taking the limit that the contraction parameter goes to infinity.

Several singular limits of geometries and/or solutions have been considered in
the recent literature for different reasons. For a few examples, see
\cite{Gomis:2005pg,Hansen:2020pqs,Fontanella:2021btt,Fontanella:2022fjd,deBoer:2023fnj}.
It is the purpose of this work to develop a general framework that describes

\begin{enumerate}
\item What happens to the Lie algebra of the isometries of the original
  metrics after the singular limits are taken.
\item What happens to the Lie algebra of symmetries of the action of a
  $p$-brane moving in the original spacetime after the singular limit is
  taken.
\end{enumerate}

(It is important to distinguish between these two symmetry algebras, because
they only need to coincide in the relativistic case.)

Our setup will be quite general. In particular we will not impose any
restrictions on the ranks of the singular metrics that arise in the limits. In
most of the studies done in this field the focus is placed on the tangent
space, which is assumed to split in a direct sum of two subspaces. The tangent
space metric is also assumed to decompose into the sum of two singular metrics
which are regular when restricted to one of the subspaces. The Vielbeins of
these two subspaces are, then, used to construct the metrics of the two
singular spacetime metrics in the standard fashion using the singular tangent
space metrics. This construction is only possible when the ranks of the
singular spacetime metrics add up to the total dimension of the spacetime, $d$
and, therefore, it cannot be used in all the cases to which our setup applies
and, in particular, it cannot be applied to the singular limit of
4-dimensional $pp$-wave metrics considered in Section~\ref{sec-pp} that
produces a metric of rank $3$ and signature $+--$\footnote{Notice that in this
  paper we use mostly minus signature throughout.} and a metric of rank $2$
and signature $+-$.

On the other hand, in order to study the symmetries of the action of
$p$-branes and its non-relativistic limits, we will make use of techniques
that have been developed within the context of non-Lorentzian limits and apply
them in a more general context.  In particular, we will consider the limit of
a sigma model action describing a $p$-brane in a generic curved background. We
will control in three different ways the leading divergence that arises when
performing in the action the redefinition that defines the limit as follows:

\vskip .2truecm

\noindent {\bf Option 1.}\ One can neutralize the leading divergence by an
appropriate rescaling of the string tension parameter and the worldvolume
metric such that the leading term scales as $\rho^{0}$ leading to a finite
answer.  We will refer to this limit as the one `with rescalings'.

\vskip.2truecm

\noindent {\bf Option 2.}\ One can tame the leading divergence by performing a
Hubbard-Stratonovich transformation which uses the fact that any term of the
form $\rho^{\alpha} X^{2} $ for some $X$ can be rewritten, by introducing an
auxiliary field $\lambda$, in the equivalent form

\begin{equation}\label{trick}
\rho^{\alpha} \big [-\frac{1}{\rho^{4}}\lambda^{2} -\frac{2}{\rho^{2}}\lambda X \big]\,.
\end{equation}

Solving for $\lambda$ and substituting this solution back one finds the
original $\rho^{\alpha} X^{2}$ term. In this work, we will apply this rewriting
for $\alpha=2$.\,\footnote{Note, however, that the same rewriting can also be
  applied for $\alpha=0$ to eliminate a finite term. For an example, see
  \cite{tobesubmitted}.} In that case, after taking the limit
$\rho \to \infty$, $\lambda$ becomes a Lagrange multiplier imposing the
constraint $X=0$. The expression  Eq.~\eqref{trick} also applies if $X$ and
$\lambda$ carry flat or curved spacetime indices.  After rewriting the
quadratic divergence according to  Eq.~\eqref{trick}, one finds a finite answer at
sub-leading order.  We will refer to this limit as the one ``with auxiliary
fields''.

\vskip .2truecm

\noindent {\bf Option 3.}\ One can cancel the leading divergence by adding a
Wess-Zumino (WZ) term to the sigma model action containing a new field and to
redefine this new field introducing a new divergence in such a way that it
neutralizes the leading divergence. A prime example of this occurs when taking
the sigma model action describing a particle coupled to general
relativity. The leading divergence can be cancelled by introducing a WZ term
containing a vector field $M_{\mu}$. After taking the limit, the non-Lorentzian
version $m_{\mu}$ of this vector field, sometimes called the mass vector field,
corresponds to a central extension of the Galilei algebra which is called the
Bargmann algebra. It describes the property that, after taking the limit,
energy and mass are two separately conserved quantities.  We will refer to
this limit as the one ``with WZ term''.

\vskip .2truecm

In this work we will not only consider the $\rho \to \infty$ limit but also
the $\rho \to 0$ limit. This option is a special case of the first one after
replacing $\rho$ by $\rho^{\prime} = 1/\rho$.  In terms of $\rho^{\prime}$ with
$\rho^{\prime} \to \infty$ there are again three ways to deal with the leading
divergence. In option 1 one can take the limit straight-away without any
rescalings since the leading ``divergence'' is already of the form
$\rho^{\prime 0}$. We will refer to this limit as one ``without rescalings''.
In option 2 one first performs a rescaling such that one obtains a leading
divergence of the form $\rho^{\prime 2}$ and next tames this divergence by
performing a Hubbard-Stratonovich transformation.  We will refer to this limit
as the one ``with auxiliary fields''.

One outcome of our general framework is that we will derive two dualities
between the Lie algebras that arise after taking singular limits with
$\rho\to\infty$ and $\rho\to 0$. One duality maps the Lie algebra that arises
after taking a singular $\rho\to\infty$ ``with auxiliary fields'' to the Lie
algebra that arises after taking a singular $\rho\to 0$ limit ``with auxiliary
fields''. Having determined the Lie algebra that arises after taking one
limit, one can apply this duality to determine the Lie algebra that arises
after taking the other limit. A prime example of this duality is the map
between the Galilei and Carroll algebras. For technical reasons, to be
explained later, we will call this duality the ``$1 \leftrightarrow -1$
duality''. This is to be contrasted with a different, formal, ``brane
duality'', which also follows from our general framework and was first noted
in \cite{Barducci:2018wuj,Bergshoeff:2020xhv}, that relates the Lie algebra
corresponding to one brane to the Lie algebra corresponding to a dual brane. A
prime example of this duality, which also follows from our general framework,
is the duality between a $p=0$ Galilei algebra and a dual $p=d-2$ Carroll
algebra. This duality is formal in the sense that one should make a formal
interchange between the single time direction of the Galilei particle and the
single spatial transverse direction of the Carroll domain wall.

This work is organized as follows. First, in Section~\ref{sec-framework}, we
will define the general sigma model action describing a $p$-brane in a special
$d$-dimensional spacetime background that contains a parameter $\rho$ whose
limit we are going to take.  We will focus on the symmetries and show how the
Killing vectors describing the symmetries before taking the limit get
transformed into the Killing vectors describing the symmetries of the foliated
geometry after taking the limit and discuss the corresponding symmetry
algebras. We will discuss both the $\rho\to\infty$ and $\rho\to 0$ limits for
two of the three options discussed above, \textit{i.e.}~by neutralizing the
leading divergence (option 1) and by performing a Hubbard-Stratonovich
transformation introducing a Lagrange multiplier (option 2). The third option
will only be applied when we discuss later in this paper the holographic limit
of an Anti-de Sitter spacetime. Next, we will apply the general framework
developed in section 2 to several examples. First, as a warming up exercise to
elucidate our techniques, we will consider in Section~\ref{sec-Minkowski} the
case that the background metric is that of a flat Minkowski spacetime. We will
do this first for particles and next generalize this to $p$-branes. In
Section~\ref{sec-holographiclimit} we will consider the more involved case of
an Anti-de Sitter (AdS) spacetime. We will consider both a so-called
holographic limit as well as a $p$-brane limit. In the case of the holographic
limit we will also consider option 3 mentioned above and add a Wess-Zumino
term involving a new 1-form gauge field to the sigma model action.  Finally,
in Section~\ref{sec-pp} we will consider a $pp$-wave background metric. We
finish with a discussion of our results and have added a separate appendix
collecting a few technical facts about AdS spacetimes that are used in the
main text.


\section{General framework}
\label{sec-framework}

In this section we construct the general framework that we will apply to
specific examples in the next sections. Our starting point is the following
$(p+1)$-dimensional $\sigma$-model (also known as the $p$-brane Polyakov-type
action):

\begin{equation}
    S\left[x^{\mu}(\zeta),\gamma_{ij}(\zeta)\right]
     =
     -\frac{T}{2}
     \int d^{p+1}\zeta \sqrt{|\gamma|}\left[\gamma^{ij}g_{ij}-(p-1)\right]\,,
     \label{eq:action}
\end{equation}
where $T$ is the tension, with units of $[ML^{-p}]$, $\zeta^{i}$, with
$i=0,1,\cdots,p$, are the worldvolume coordinates, $\gamma_{ij}(\zeta)$ is the
worldvolume metric, $|\gamma|$ is the absolute value of its determinant and
$\gamma^{ij}$ its inverse, $g_{\mu\nu}(x)$ is the $d$-dimensional spacetime
metric and $g_{ij} = \partial_{i}x^{\mu} \partial_{j}x^{\nu} g_{\mu\nu}$, is
its pullback to the worldvolume.

We are going to consider spacetime metrics that can be written in the form

\begin{equation}
  \label{eq:assumptionmetric}
  g_{\mu\nu}
  =
  h_{\mu\nu}+\rho^{2}  k_{\mu\nu}\,,
\end{equation}

\noindent
where the ranks of the metrics $h_{\mu\nu}$ and $k_{\mu\nu}$ (assumed to be
finite) is smaller than $d$ and $\rho$ is a parameter whose limit we will take
to zero and infinity.\footnote{Although it seems natural to expect that the
  sums of the ranks of these metrics is $d$, as we have explained in the
  introduction, we are not going to make this assumption and, as a matter of
  fact, in Section~\ref{sec-pp} we are going to study an example in which the
  sum is $d+1$. This is perfectly consistent in our setup, since the metric is
  a background field with no degrees of freedom attached but one has to bear
  in mind that, in a different context, this may turn out to be inconsistent.}
In these two limits the spacetime metric $g_{\mu\nu}$ becomes singular or
ill-defined and must be replaced by the singular but well-defined, metrics
$h_{\mu\nu}$ and $k_{\mu\nu}$ in terms of which the action  Eq.~\eqref{eq:action}
takes the equivalent form

\begin{equation}
  \label{eq:actionintermsofsingularmetrics}
    S\left[x^{\mu}(\zeta),\gamma_{ij}(\zeta)\right]
     =
     -\frac{T}{2}
     \int d^{p+1}\zeta \sqrt{|\gamma|}\left[\gamma^{ij}h_{ij}
       +\rho^{2}\gamma^{ij}k_{ij}-(p-1)\right]\,.
\end{equation}

We want to study the symmetries of this action (specially the global ones)
before and after taking the limits in which  $\rho$ goes to zero or infinity. 

Before taking the limits, the global symmetries of the action
\eqref{eq:actionintermsofsingularmetrics} are generated by infinitesimal
transformations of the embedding coordinates $x^{\mu}(\zeta)$ of the form

\begin{equation}
\delta x^{\mu} = \epsilon\xi^{\mu}(x)\,,
\end{equation}

\noindent
where $\epsilon$ is an infinitesimal (constant) parameter and $\xi^{\mu}$ is a
Killing vector of the metric $g_{\mu\nu}$, that is

\begin{equation}
  \pounds_{\xi}g_{\mu\nu}
  =
  0\,.
\end{equation}

\noindent
Thus, effectively, the study of the symmetries of the action is equivalent to
the study of the isometries of the metric, at least before taking the limits.
It is natural, then, to start our investigation by studying the Killing
vectors of the original metric and the Lie algebra they generate.

\subsection{The Lie algebra of isometries}
\label{sec-Liealgebraofisometries}

We are going to assume that all the Killing vectors admit an expansion in the
parameter $\rho$ of the form

\begin{equation}\label{expansion}
\xi = \frac{1}{\rho}\xi^{(-1)}  +\xi^{(0)} +\rho\xi^{(1)}\,,
\end{equation}

\noindent
and, furthermore, that these Killing vectors generate a Lie algebra with
$\rho$-independent structure constants. This assumption is backed by the fact
that if we add terms of higher or lower orders in $\rho$ to the expansion
\eqref{expansion} they will correspond to Killing vectors which are already
included in the above expansion, multiplied by powers of $\rho$. It is also
backed by the examples that we are going to present in the following sections.

Observe that the components in the above expansion,
$\xi^{(-1)},\xi^{(0)},\xi^{(1)}$, are (non-necessarily Killing) vector fields
whose Lie brackets we can always compute. We can also compute the Lie
derivatives of the singular metrics with respect to each of them.

Using the expansion  Eq.~\eqref{expansion} and the main hypothesis about the
spacetime metric Eq.~(\ref{eq:assumptionmetric}), we find at each order in
$\rho$ that the following $\rho$-independent equations must be satisfied:

\begin{subequations}
  \begin{align}
    \pounds_{\xi^{(1)}}k_{\mu\nu}
    & =
      0\,,
    \\
    & \nonumber \\
    \pounds_{\xi^{(0)}}k_{\mu\nu}
    & =
      0\,,
    \\
    & \nonumber \\
          \pounds_{\xi^{(-1)}}k_{\mu\nu}
          +\pounds_{\xi^{(1)}}h_{\mu\nu}
    & =
      0\,,
    \\
    & \nonumber \\
    \pounds_{\xi^{(0)}}h_{\mu\nu}
    & =
      0\,,
    \\
    & \nonumber \\
    \pounds_{\xi^{(-1)}}h_{\mu\nu}
    & =
      0\,.
  \end{align}
\end{subequations}

That is: the $\xi^{(0)}$ vector fields are Killing vectors of the two singular
metrics and, therefore, they are Killing vectors of the original metric. The
vector fields $\xi^{(1)}$ (resp.~-1) are Killing vectors of the singular
metric $k_{\mu\nu}$ (resp.~$h_{\mu\nu}$) only.  The isometry algebra of the
singular metric $k_{\mu\nu}$ (resp.~$h_{\mu\nu}$) is the Lie algebras
generated by the vectors $\xi^{(0)}$ and $\xi^{(1)}$ (resp.~$\xi^{(0)}$ and
$\xi^{(-1)}$), which must be necessarily closed (see the discussion below.)

At first sight one may conclude that, if both singular metrics survive in the
action after the limit, the algebra of symmetries of the action should be one
generated by the the $\xi^{(0)}$ vector fields, which should, therefore,
generate a closed subalgebra of the original isometry algebra. If only
$k_{\mu\nu}$ (resp.~$h_{\mu\nu}$) survived in the action, the algebra of
symmetries would be the one generated by the vectors $\xi^{(0)}$ and
$\xi^{(1)}$ (resp.~$\xi^{(0)}$ and $\xi^{(-1)}$). As we are going to see,
though, this conclusion is wrong: the one-to-one relationship between Killing
vectors of the metric that occurs in the action and the generators of
symmetries of the action breaks down in the limits.\footnote{This is a well
  known fact that, perhaps, has not been formulated before in this way. The
  symmetry group of the action of the non-relativistic particle obtained in
  the $c\rightarrow \infty$ limit of the standard relativistic particle
  \begin{equation}
    S[t,x^{i}]
    \sim
    \int d \zeta \frac{\delta_{ij}\dot{x}^{i}\dot{x}^{j}}{\dot{t}}\,,    
  \end{equation}
  (the Galileo group) has the same dimension as the Poincar\'e group even
  though only the Euclidean metric $\delta_{ij}$, and a trivial 1-dimensional
  metric in the time direction occur in it.} The missed ingredient in this
analogy is the behaviour of the transformation parameters, which can be
rescaled with powers of $\rho$ so as to absorb divergences. Using these
rescalings one can obtain a symmetry of the action for each of the original
symmetries/isometries \textit{after} taking the limits. We will see how this
works out in the following sections, but we first need to study in more detail
the brackets of all the vector fields involved.

Our assumptions for the expansion of the Killing vectors in powers of $\rho$
and the $\rho$-independence of the structure constants have the following
consequences for the Lie algebra:

\begin{enumerate}
\item The following Lie brackets vanish

  \begin{equation}
    [\xi_{A}^{(1)},\xi_{B}^{(1)}]
    =
    [\xi_{A}^{(-1)},\xi_{B}^{(-1)}]
    =
    0\,,
  \end{equation}
  where the index $A$ labels the different Killing vectors of which
  $\xi_{A}^{(1)}$ and $\xi_{A}^{(-1)}$ are components.

\item The Lie brackets of $\xi^{(0)}$ vectors close on $\xi^{(0)}$ vectors
  and, hence, they generate a subalgebra. Since they are Killing vectors of
  $h$ and $k$ simultaneously, one can always find a basis of the Lie algebra
  in which there are just two kinds of vectors: those which are independent of
  $\rho$, so $\xi=\xi^{(0)}\equiv \pi$ and those which do not contain any
  $\rho$-independent piece and are of the form
  $\xi=\rho^{-1}\xi^{(-1)}+\rho\xi^{(1)} \equiv \varpi$, where either
  $\xi^{(-1)}$ or $\xi^{(1)}$ may vanish.  Henceforth we will always use this
  basis.

\item The Lie brackets of $\pi$ vectors and $\varpi$ vectors close on
  $\varpi$ vectors only. Thus, the $\varpi$ vectors span a representation of
  the subalgebra of the $\pi$ vectors.

\item The Lie bracket of two $\varpi$ vectors is always a $\pi$ vector.

\end{enumerate}

The last three statements can be summarized symbolically by the following Lie
brackets:

\begin{equation}
  \label{eq:Liealgebrapattern}
  [\pi,\pi] = \pi\,,
  \hspace{1cm}
  [\pi,\varpi] = \varpi\,,
  \hspace{1cm}
  [\varpi,\varpi] = \pi\,,
\end{equation}

\noindent
which describe a symmetric decomposition of the Lie algebra of the original
isometry group of $g_{\mu\nu}$. Furthermore, all the $\varpi$ vectors are of
the form

\begin{equation}
\varpi = \frac{1}{\rho}\varpi^{(-1)} +\rho\varpi^{(1)}\,,
\end{equation}

\noindent
and the only non-trivial Lie brackets between the $\varpi $ components are

\begin{equation}
[\varpi^{(-1)},\varpi^{(1)}]  =\pi\,,
\end{equation}

\noindent
and between the $\varpi$ and $\pi$ components are\,\footnote{ As we have
  remarked before, in general, these Lie brackets are not part of the Lie
  algebra because, in general, only the combination
  $\rho^{-1}\varpi^{(-1)} +\rho\varpi^{(1)}$ is actually a
  generator. Nevertheless, these Lie brackets are important because, depending
  on the limit taken, either $\varpi^{(-1)}$ or $\varpi^{(1)}$ will disappear
  and the surviving component will become a generator of the new
  (Wigner-In\"on\"u-contracted) algebra of symmetries of our system even if
  they are not Killing vectors anymore.}

\begin{equation}
[\pi,\varpi^{(-1)}]  =\varpi^{(-1)}\,,
\,\,\,\,
\text{and}
\,\,\,\,
[\pi,\varpi^{(1)}]  =\varpi^{(1)}\,.
\end{equation}

\noindent
This means that we are free to rescale

\begin{equation}
  \varpi^{(-1)} \rightarrow \rho^{-n}\varpi^{(-1)}\,,
  \hspace{1cm}
  \varpi^{(1)} \rightarrow \rho^{n}\varpi^{(1)}\,,
\end{equation}

\noindent
preserving the structure constants of the group. In particular, we could
rescale with $n=-1$ so that

\begin{equation}
\varpi =   \varpi^{(-1)} +\rho^{2}\varpi^{(1)}\,,
\end{equation}

\noindent
which is a more standard expansion.

There are two special cases that need to be considered: when either
$\varpi^{(1)}=0$ or $\varpi^{(-1)}=0$, the other non-zero component in the
expansion ($\varpi^{(-1)}$ or $\varpi^{(1)}$, respectively) is a Killing
vector of the whole metric and of each of the two singular metrics. These
Killing vectors will always survive the limits because their dependence on
$\rho$ can always be removed and, therefore, they need to be taken into
account.

We are now ready to study the limits of the sigma model action
\eqref{eq:actionintermsofsingularmetrics} and of its symmetries. We can take
four different limits (there are two different ways of taking the
$\rho\rightarrow 0,\infty$ limits) and we consider them separately in the
following sections.

\subsection{$\rho\rightarrow\infty$
  with auxiliary fields}
\label{sec-rhotoinftywithauxiliaryfields}

Performing a Hubbard-Stratonovich transformation Eq.~\eqref{trick} for
$\alpha=2$, we rewrite the action
Eq.~\eqref{eq:actionintermsofsingularmetrics} in the following form

\begin{equation}\label{actionaf}
    S\left[x^{\mu}(\zeta),\gamma_{ij}(\zeta)\right]
     =
     -\frac{T}{2}
     \int d^{p+1}\zeta \sqrt{|\gamma|}\left[\gamma^{ij}h_{ij}
       -\varepsilon\left(\frac{1}{4\rho^{2}}\lambda^{2}
         +\lambda \sqrt{|\gamma^{ij}k_{ij}|}\right)-(p-1)\right]\,,
\end{equation}

\noindent
where we have introduced a parameter $\varepsilon = \pm 1$ with
$\varepsilon=+1$ if the time coordinate is included in $k$ and where
$\varepsilon=-1$ otherwise. In this way we always have\,\footnote{If the time
  direction is included in $k$ (that is: if one of the signs of the signature
  of $k$ is a $+$), then there should always be a coordinate system in which
  only the time coordinate is non-constant and, then $\gamma^{ij}k_{ij}>0$
  with our mostly minus signature convention for the metric. For point
  particles, this is equivalent to the assumption that the worldlines should
  always be timelike which is implicitly made in the Nambu-Goto- and
  Polyakov-type actions.}

\begin{equation}
  \varepsilon |\gamma^{ij}k_{ij}|
  =
  \gamma^{ij}k_{ij}\,.
\end{equation}

The symmetries of the action  Eq.~\eqref{actionaf} are preserved provided $\lambda$
transforms as the solution of its equation, namely
$\lambda = -2\rho^{2} \sqrt{|\gamma^{ij}k_{ij}|}$. Thus,

\begin{equation}
  \begin{aligned}
    \delta\lambda
    & =
    \rho \frac{\varepsilon}{\sqrt{|\gamma^{ij} k_{ij}|}}
    \gamma^{ij}\partial_{i}x^{\mu}\partial_{j}x^{\nu}
    \pounds_{\xi^{(-1)}} k_{\mu\nu}\,.
  \end{aligned}
\end{equation}

Observe that, in the special case in which $\xi^{(1)}=0$ so that $\xi^{(-1)}$
is a Killing vector of $g,h$ and $k$, this variation vanishes automatically,
$\lambda$ is invariant and so is the full action.

It is important to realize that in the present case we cannot speak properly
of isometries and Killing vectors because the variable $\lambda$ is not a
coordinate nor the derivative of a coordinate and has no known geometrical
interpretation.  We can still speak of the Lie algebra of symmetries of this
action, though, and it is clear that all of them\footnote{Strictly speaking
  one can exclude additional emergent symmetries unrelated to the isometries
  of the original metric.} are associated to the Killing vectors of the
original metric and their Lie brackets correspond to the commutators of the
generators of the symmetries of the action.

Now we introduce infinitesimal transformation parameters $a^{m},b^{x}$ for
every Killing vector $\pi_{m},\varpi_{x}$, defining the following
transformations of the variables of the action

\begin{equation}
  \begin{aligned}
    \delta x^{\mu}
    & =
    a^{m}\pi_{m}^{\mu}
    +\frac{b^{x}}{\rho}\varpi_{x}^{(-1)\,\mu}+b^{x}\rho\,\varpi_{x}^{(1)\,\mu}\,,
    \\
    & \\
    \delta\lambda
    & =
       b^{x} \rho \frac{\varepsilon}{\sqrt{|\gamma^{ij} k_{ij}|}}
    \gamma^{ij}\partial_{i}x^{\mu}\partial_{j}x^{\nu}
    \pounds_{\varpi_{x}^{(1)}} h_{\mu\nu}\,.
  \end{aligned}
\end{equation}

Assuming that $\varpi_{x}^{(1)} \neq 0$, in order to get a finite result in
the $\rho\rightarrow \infty$ limit of these transformations, we must replace

\begin{equation}
b^{x} \rightarrow \frac{b^{x}}{\rho}\,.
\end{equation}

After taking the $\rho\to\infty$ limit, we end up with  the following action:

\begin{equation}
  \label{eq:roinftylimitnorescalingsaction}
    S\left[x^{\mu}(\zeta),\gamma_{ij}(\zeta)\right]
     =
     -\frac{T}{2}
     \int d^{p+1}\zeta \sqrt{|\gamma|}\left[\gamma^{ij}h_{ij}
     -\varepsilon\lambda \sqrt{|\gamma^{ij}k_{ij}|}-(p-1)\right]\,.
\end{equation}
which is invariant under the transformation rules

\begin{equation}
  \begin{aligned}
    \delta x^{\mu}
    & =
    a^{m}\pi_{m}^{\mu} +b^{x}\varpi_{x}^{(1)\,\mu}\,,
    \\
    & \\
    \delta\lambda
    & =
       b^{x}\frac{\varepsilon}{\sqrt{|\gamma^{ij} k_{ij}|}}
    \gamma^{ij}\partial_{i}x^{\mu}\partial_{j}x^{\nu}
    \pounds_{\varpi_{x}^{(1)}} h_{\mu\nu}\,.
  \end{aligned}
\end{equation}

Notice that there is one symmetry generator for each of the isometry
generators of the original metric and each of the symmetries of the original
action.

This is clearly correct when the components $\varpi_{x}^{(1)}\neq 0$. When
they vanish, we can make the opposite rescaling $b^{x}\rightarrow b^{x}\rho$
and the symmetry generated by $\varpi_{x}^{(-1)}$ will survive the limit.

It is convenient to split the index $x$ as follows: $x_{-1}$ which corresponds
to $\varpi$ generators with no $\varpi^{(1)}$ component, $x_{1}$ which
corresponds to $\varpi$ generators with no $\varpi^{(-1)}$ component, and
$x_{2}$, which corresponds to $\varpi$ generators with both components:

\begin{equation}
  \{\varpi_{x}\}
  =
  \{\varpi_{x_{-1}},\varpi_{x_{1}},\varpi_{x_{2}}\}\,,
\end{equation}

\noindent
By definition, they are of the form

\begin{equation}
  \varpi_{x_{-1}}
  =
  \frac{1}{\rho}\varpi_{x_{-1}}^{(-1)}\,,
  \hspace{1cm}
  \varpi_{x_{1}}
  =
  \rho\varpi_{x_{1}}^{(1)}\,,
  \hspace{1cm}
  \varpi_{x_{2}}
  =
  \frac{1}{\rho}\varpi_{x_{2}}^{(-1)}
  +
  \rho\varpi_{x_{2}}^{(1)}\,.
\end{equation}

Using this notation, we only have to take into account the $ \varpi_{x_{-1}}$
components at the very end, to compute the final symmetry algebra.

The algebra of symmetries is generated by

\begin{enumerate}
\item The vectors $\pi_{m}$, which were Killing vectors of the original metric
  $g$ and remain Killing vectors of the two degenerate metrics $h$ and $k$.
\item The vectors $\varpi_{x}^{(1)}$ coming from the
  $\varpi_{x_{1}}^{(1)},\varpi_{x_{2}}^{(1)}$, which are Killing vectors of
  the degenerate metric $k$ only. Their index $x$ runs over all the values.
\item The vectors $\varpi_{x_{-1}}^{(-1)}$ which are Killing vectors of $g,h$
  and $k$.

\end{enumerate}

The algebra of symmetries is determined by the non-vanishing brackets

\begin{align}
  \label{eq:therhoinfinityalgebra}
    [\pi_{m},\pi_{n}] = f_{mn}{}^{p}\pi_{p}\,, \hspace{1cm} &
    [\pi_{m},\varpi^{(1)}_{x}] = f_{mx}{}^{y}\varpi^{(1)}_{y}\,,
    \\
    & \nonumber \\
    [\pi_{m},\varpi^{(-1)}_{x_{-1}}] =
    f_{mx_{-1}}{}^{y_{-1}}\varpi^{(-1)}_{y_{-1}}\,, \hspace{1cm} &
    [\varpi^{(-1)}_{x_{-1}},\varpi^{(1)}_{y}] = f_{x_{-1} y}{}^{m}\pi_{m}\,.
\end{align}

Although the specific details will differ, some patterns we have seen in this
first limit (in particular, the rescaling of the transformation parameters and
the survival after the limit of the same number of symmetries present in the
original action, with some different commutators) will repeat themselves in
the other three limits that we are going to study next.

\subsection{$\rho\rightarrow 0$ with auxiliary fields}
\label{sec-rhozerolimits}

The $\rho\rightarrow 0$ limits with auxiliary fields are closely related to
the $\rho \to \infty$ limits with auxiliary fields discussed
above. Effectively, after replacing $\rho$ by $1/\rho$ and a suitable
rescaling, the two actions are related by an interchange of the degenerate
metrics $h_{\mu\nu}$ and $k_{\mu\nu}$. It has, however, the effect, that one
obtains a slightly different algebra of symmetries in this limit. One finds
that the transformations of the coordinates after the limit is taken (with the
necessary rescalings of the parameters) are given by

\begin{equation}
    \delta x^{\mu}
    =
    a^{m}\pi_{m}^{\mu} +b^{x}\varpi_{x}^{(-1)\,\mu}
    +b^{x_{1}}\varpi_{x_{1}}^{(1)\,\mu}\,.
  \end{equation}

\noindent
These symmetries are generated by

\begin{enumerate}
\item The vectors $\pi_{m}$, which were Killing vectors of the original metric
  $g$ and remain Killing vectors of the two degenerate metrics $h$ and $k$.
\item The vectors $\varpi_{x}^{(-1)}$ coming from the
  $\varpi_{x_{-1}}^{(-1)},\varpi_{x_{2}}^{(-1)}$, which are Killing vectors of
  the degenerate metric $h$ only. Their index $x$ runs over all the values.
\item The vectors $\varpi_{x_{1}}^{(1)}$ which are Killing vectors of $g,h$
  and $k$.

\end{enumerate}

The algebra of symmetries takes the form

\begin{align}
  \label{eq:therhozeroalgebra}
      [\pi_{m},\pi_{n}] = f_{mn}{}^{p}\pi_{p}\,, \hspace{1cm} &
    [\pi_{m},\varpi^{(-1)}_{x}] = f'_{mx}{}^{y}\varpi^{(-1)}_{y}\,,
    \\
    & \nonumber \\
    [\pi_{m},\varpi_{x_{1}}^{(1)}] = f'_{mn}{}^{x_{1}}\varpi_{x_{1}}^{(1)}\,,
    \hspace{1cm} &
    [\varpi^{(-1)}_{x},\varpi_{x_{1}}^{(1)}] = f_{xx_{1}}{}^{m}\pi_{m}\,.
\end{align}

It is worth comparing this algebra with the algebra obtained in the
$\rho\rightarrow\infty$ limit, given in
Eq.~(\ref{eq:therhoinfinityalgebra}). The structure is very similar: one can
be formally obtained from the other by exchanging all the $1$ and $-1$
indices, except for those generators which have $x_{1},x_{-1}$ subindices,
which survive both limits. It is worth stressing that the final structure
constants $f'_{mx}{}^{y}$ are, in general, different from the final
$f_{mx}{}^{y}$ structure constants obtained in the $\rho\rightarrow \infty $
limit and that neither of them has to coincide with the original structure
constants in the $[\pi,\varpi]$ brackets. This can be seen explicitly in the
examples (see, for instance, Tables~\ref{figure1} and~\ref{tab:fig2}).

\subsection{$\rho\rightarrow\infty$ with rescalings}

There is a second procedure for taking the $\rho\rightarrow \infty$ limit
which does not require the use of auxiliary fields: removing the leading
divergence by rescaling the worldvolume metric and the tension according to

\begin{equation}
  \gamma_{ij}\rightarrow \rho^{2}\gamma_{ij}\,,
  \hspace{1cm}
  T\rightarrow \rho^{-(p+1)}T\,,
\end{equation}

\noindent
after which the action  Eq.~\eqref{eq:action} takes the form

\begin{equation}
    S\left[x^{\mu}(\zeta),\gamma_{ij}(\zeta)\right]
     =
     -\frac{T}{2}
     \int d^{p+1}\zeta \sqrt{|\gamma|}\left[\frac{1}{\rho^{2}}\gamma^{ij}h_{ij}
       +\gamma^{ij}k_{ij}-(p-1)\right]\,,
\end{equation}

\noindent
We can now take the $\rho\rightarrow \infty$ limit directly without
introducing an auxiliary field, obtaining

\begin{equation}
  \label{eq:rhoinfinitylimitrescalingsaction}
    S\left[x^{\mu}(\zeta),\gamma_{ij}(\zeta)\right]
     =
     -\frac{T}{2}
     \int d^{p+1}\zeta \sqrt{|\gamma|}\left[\gamma^{ij}k_{ij}-(p-1)\right]\,.
\end{equation}

Taking the limit in the transformation rules, we obtain the same rules as
before:

\begin{equation}
    \delta x^{\mu}
    =
    a^{m}\pi_{m}^{\mu} +b^{x_{-1}}\varpi_{x_{-1}}^{(-1)\,\mu}
    +b^{x}\varpi_{x}^{(1)\,\mu}\,.
\end{equation}

\noindent
These symmetries are generated by the vectors
$\pi_{m},\varpi_{x_{-1}}^{(-1)}$, which are Killing vectors of $g,h$ and $k$
and by the vectors $\varpi_{x}^{(1)}$ which are Killing vectors of $k$
only. The invariance of the action under these symmetries follows trivially.
Furthermore, the algebra of symmetries is the same as the one we found before
in  Eq.~\eqref{eq:therhoinfinityalgebra}.

\subsection{$\rho\rightarrow 0$ without rescalings}

The $\rho\rightarrow 0$ limit the action
Eq.~(\ref{eq:actionintermsofsingularmetrics}) can be taken without rescalings
or auxiliary variables.  We get directly
Eq.~(\ref{eq:rhoinfinitylimitrescalingsaction}) with the degenerate metric
$k_{\mu\nu}$ replaced by $h_{\mu\nu}$ and the same symmetry algebra we just
found above.

It is worth mentioning that the action of some of the symmetry generators on
the actions obtained by taking limits without the use of auxiliary fields can
be trivial.

\vskip .2truecm

Having explained the general techniques, we are now going to apply them in the
next sections to several examples.

\section{Example 1: The standard limits of Minkowski spacetime}
\label{sec-Minkowski}

In this section we will recover some very well-known results for Minkowski
spacetime using the machinery developed in the previous section, putting it to
the test. We will first consider particle limits with $p=0$ and then we will
extend these limits to $p$-brane limits where $p$ is general.

\subsection{Particle limits}
\label{sec-Minkowskiparticle}

Our starting point is the  Minkowski metric in Cartesian coordinates  given by

\begin{equation}
  ds^{2}
  =
  c^{2}dt^{2} -dx^{m}dx^{m}\,,
  \hspace{1cm}
  m=1,\cdots, d-1\,.
\end{equation}

\begin{equation}
  h_{\mu\nu}dx^{\mu}dx^{\nu}
  =
 -dx^{m}dx^{m}\,,
  \hspace{1cm}
  k_{\mu\nu}dx^{\mu}dx^{\nu}
  =
  dt^{2}\,,
    \hspace{1cm}
    \rho
    =
    c\,.
\end{equation}
The Killing vectors are given by

\begin{equation}
  \label{eq:KillingvectorsMinkowskiCartesian}
  \xi_{\mu}
  =
  \partial_{\mu}\,,
  \hspace{1cm}
  \xi_{\mu\nu}
  =
  2\eta_{\mu\nu}{}^{\rho}{}_{\sigma}x^{\sigma}\partial_{\rho}\,,
\end{equation}

\noindent
and satisfy the Lie algebra

\begin{subequations}
  \label{eq:LieAlgebraKillingvectorsMinkowskiCartesian}
  \begin{align}
    [\xi_{\mu\nu},\xi_{\rho\sigma}]
    & =
    +\eta_{\mu\rho}\xi_{\nu\sigma}
    +\eta_{\nu\sigma}\xi_{\mu\rho}
    -\eta_{\mu\sigma}\xi_{\nu\rho}
    -\eta_{\nu\rho}\xi_{\mu\sigma}\,,
    \\
    & \nonumber \\
    [\xi_{\mu},\xi_{\rho\sigma}]
    & =
      -2\eta_{\mu[\rho}\xi_{\sigma]}\,.
  \end{align}
\end{subequations}
They admit an expansion in the parameter $c$ so that they can be identified with the vectors in the general framework in the following way

\begin{subequations}
  \begin{align}
  \xi_{0}
  & =\varpi_{0}=
    \frac{1}{c}\partial_{t}\,,
    \\
    & \nonumber \\
  \xi_{m}
  & = \pi_{m}=
    \partial_{m}\,,
    \\
    & \nonumber \\
  \xi_{m 0}
  & =\varpi_{m 0}=
ct\partial_{m} +\frac{1}{c}x^{m}\partial_{t}\,,
    \\
    & \nonumber \\
  \xi_{m n}
    & = \pi_{m n}=
      x^{m}\partial_{n}-x^{n}\partial_{m}\,.
  \end{align}
\end{subequations}

\noindent

The non-vanishing  Lie brackets of these Killing vectors are\,\footnote{Remember
that our Minkowski metric has mostly minus signature.}

\begin{subequations}
  \begin{align}
   [\pi_{mn}, \pi_{pq}]
    & =
    -\delta_{mp}\pi_{nq}
    -\delta_{nq}\pi_{mp}
    +\delta_{mq}\pi_{np}
    +\delta_{np}\pi_{mq}\,,
    \\
    & \nonumber \\
   [\pi_{m}, \pi_{np}]
    & =
2\delta_{m[n}\pi_{p]}\,,
    \\
    & \nonumber \\
   [ \pi_{mn},\varpi_{p 0}]
    & =
      -2\delta_{p[m}\varpi_{n]0}\,,
      \\
    & \nonumber \\
   [\pi_{m},\varpi_{n 0}]
    & =
      \delta_{mn}\varpi_{0}\,,
      \\
    & \nonumber \\
    [\varpi_{0},\varpi_{m 0}]
    & =
      \pi_{m}\,,
    \\
    & \nonumber \\
    [\varpi_{(m0)},\varpi_{(n0)}]
    & =
     \pi_{mn}\,,
  \end{align}
\end{subequations}

\noindent
which fits  the general pattern of  Eq.~(\ref{eq:Liealgebrapattern}).
Furthermore, we see that the subalgebra generated by the $\pi$s (which is the
algebra of the Euclidean group in $d-1$ dimensions) leaves invariant the two
singular metrics

\begin{equation}
  \pounds_{\pi}k_{\mu\nu}
  =
  \pounds_{\pi}h_{\mu\nu}
  =
  0\,,
\end{equation}

\noindent
Defining

\begin{subequations}
  \begin{align}
  \varpi^{(-1)}_{0}
  & \equiv
    \partial_{t}\,,
    \\
    & \nonumber \\
  \varpi^{(-1)}_{m 0}
  & \equiv
x^{m}\partial_{t}\,,
    \\
    & \nonumber \\
  \varpi^{(1)}_{m 0}
  & \equiv
t\partial_{m}\,,
  \end{align}
\end{subequations}

\noindent
we find

\begin{equation}
  \pounds_{\varpi^{(-1)}}h_{\mu\nu}
  =
  \pounds_{\varpi^{(1)}}k_{\mu\nu}
  =
  0\,,
\end{equation}

\noindent
and

\begin{equation}
  \pounds_{\varpi^{(1)}_{m0}}h_{\mu\nu}
  +
  \pounds_{\varpi^{(-1)}_{m0}}k_{\mu\nu}
  =
  0\,,
  \hspace{1cm}
  \pounds_{\varpi^{(-1)}_{0}}k_{\mu\nu}
  =
  0\,.
\end{equation}

The Killing vector $\varpi^{(-1)}_{0}$ belongs to the class of those which do
not have a $\varpi^{(1)}$ component and generate symmetries of $g,h$ and $k$. It will therefore
 survive all the limits.

\subsubsection{$c\rightarrow \infty$
with auxiliary fields}

The $c\rightarrow\infty$ limit of the $p$-brane action in Minkowski spacetime
takes the general form in Eq.~(\ref{eq:roinftylimitnorescalingsaction}) with $\epsilon=1$. For
the particular case $p=0$ in which the only component of the worldline metric
$\gamma$ is replaced by the Einbein $e$ ($\gamma=e^{2}$) and the constant $T$ takes the form
$T=mc \rightarrow \tilde{m}$, we get

\begin{equation}
  \begin{aligned}
    S\left[x^{\mu}(\zeta),e(\zeta)\right]
    & =
    \frac{\tilde{m}}{2}
    \int d\zeta \left[e^{-1}\dot{x}^{m}\dot{x}^{m}
      +\lambda \dot{t} -e\right]\,.
  \end{aligned}
\end{equation}
This action is invariant under the global transformations

\begin{equation}
  \delta x^{\mu}
  =
  a^{m}\pi_{m}{}^{\mu}+\tfrac{1}{2}\sigma^{mn}\pi_{mn}{}^{\mu}
  +b^{m}\varpi^{(1)}_{m0}+c\varpi^{(-1)}_{0}\,.
\end{equation}
The algebra of symmetries of this action is that generated by the
$\{\pi_{m},\pi_{mn}\}$ generators (the Euclidean group in $d-1$ dimensions)
together with the $\varpi^{(-1)}_{0}$ and $\varpi^{(1)}_{m0}$ generators,
which have the following non-trivial Lie brackets:

\begin{equation}
\label{Galilei}
  [\pi_{mn},\varpi^{(1)}_{p0}]
  =
  -2\delta_{p[m}\varpi^{(1)}_{n]0}\,,
  \hspace{1cm}
  [\varpi^{(-1)}_{0},\varpi^{(1)}_{m0}]
  =
  \pi_{m}\,,
\end{equation}

This is the algebra of the \textit{Galilei group}.

\bigskip

The equation of motion of the Lagrange multiplier $\lambda$ is $\dot{t}=0$ and, therefore, we can eliminate the
$\lambda \dot{t}$ term from the action. The solution to the equation of motion
of the Einbein is

\begin{equation}
e =\sqrt{-\dot{x}^{m}\dot{x}^{m}}\,.
\end{equation}
Following \cite{Batlle:2017cfa}, the problematic negative sign of the square root argument problem can be fixed by starting with a tachyonic particle replacing
$\tilde{m}\rightarrow i\tilde{m}$.\footnote{For a tachyonic particle,
  $g_{\mu\nu}\dot{x}^{\mu}\dot{x}^{\nu}<1$ in our conventions and
  $\sqrt{g_{\mu\nu}\dot{x}^{\mu}\dot{x}^{\nu}}$ is imaginary. Therefore  the constant
  in front of the action must also be imaginary in order for the  the action to be real.}
After that change we arrive at the action of a massless Galilean particle

\begin{equation}
    S\left[x^{m}(\zeta)\right]
    =
    \tilde{m} \int d\zeta \sqrt{\dot{x}^{m}\dot{x}^{m}}\,.
\end{equation}

\subsubsection{$c\rightarrow 0$
with auxiliary fields}

As indicated in Section~\ref{sec-rhozerolimits}, the action for this case can
be obtained by interchanging the degenerate metrics $h$ and $k$ in the
original action, while at the same time replacing $c$ by $1/c$. For $p=0$,
$Tc=mc^{2} \rightarrow \tilde{m}$ and $\varepsilon=-1$, the resulting action
is given by \cite{Bergshoeff:2014jla}

\begin{equation}
  \begin{aligned}
    S\left[x^{\mu}(\zeta),e(\zeta)\right]
    & =
    -\frac{\tilde{m}}{2} \int d\zeta \left[e^{-1}\dot{t}^{2}
      +\lambda \sqrt{\dot{x}^{m}\dot{x}^{m}}+e\right]\, ,
  \end{aligned}
\end{equation}
which is invariant under the global transformations

\begin{equation}
  \delta x^{\mu}
  =
  a^{m}\pi_{m}{}^{\mu}+\tfrac{1}{2}\sigma^{mn}\pi_{mn}{}^{\mu}
  +b^{m}\varpi^{(-1)}_{m0}+c\varpi^{(-1)}_{0}\,.
\end{equation}

\bigskip

The algebra of symmetries is that generated by the $\{\pi_{m},\pi_{mn}\}$
generators (the Euclidean group in $d-1$ dimensions) together with the
$\{\varpi^{(-1)}_{0},\varpi^{(-1)}_{m0}\}$ generators, which only have the
following non-trivial Lie brackets with the $\pi$s:

\begin{equation}
\label{Carrol}
  [\pi_{mn},\varpi^{(-1)}_{p0}]
  =
  -2\delta_{p[m}\varpi^{(-1)}_{n]0}\,,
  \hspace{1cm}
  [\pi_{m},\varpi^{(-1)}_{n0}]
  =
  \delta_{mn}\varpi^{(-1)}_{0}\,.
\end{equation}
This is the algebra of the \textit{Carroll group}.

The equation of motion of $\lambda$ eliminates any possible dynamics:
$\dot{x}^{m}=0$. The remaining terms in the action do not contain any degrees
of freedom.

\subsubsection{$c\rightarrow \infty$ with rescalings}

The $c\rightarrow \infty$ limit can also be taken by rescaling the Einbein $e$ and $\tilde{m}$  thus removing the leading divergence in the metric in the following way

\begin{equation}
   \tilde{m}\rightarrow c^{-1} \tilde{m} \,,
  \hspace{1cm}
 e \rightarrow c^{2} e\,.
\end{equation}

The particle action takes the form

\begin{equation}
  \begin{aligned}
    S\left[x^{\mu}(\zeta),e(\zeta)\right]
    & =
  -  \frac{\tilde{m}}{2} \int d\zeta \left[\frac{1}{c^{2}}e^{-1}\dot{x}^{m}\dot{x}^{m}+e^{-1}\dot{t}^{2}
      +e\right]\, ,
  \end{aligned}
\end{equation}

We can take the $c\rightarrow \infty$ limit without introducing any auxiliary
field, obtaining

\begin{equation}
  \begin{aligned}
    S\left[x^{\mu}(\zeta),e(\zeta)\right]
    & =
    -\frac{\tilde{m}}{2} \int d\zeta \left[e^{-1}\dot{t}^{2}
      +e\right]\, ,
  \end{aligned}
\end{equation}

The limit in the transformation laws leads to

\begin{equation}
  \delta x^{\mu}
  =
  a^{m}\pi_{m}{}^{\mu}+\tfrac{1}{2}\sigma^{mn}\pi_{mn}{}^{\mu}
  +b^{m}\varpi^{(1)}_{m0}+c\varpi^{(-1)}_{0}\, ,
\end{equation}
and the algebra of the symmetries is the one of the Galilei group  Eq.~\eqref{Galilei}.

\subsubsection{$c\rightarrow 0$ without rescalings}

The $c\rightarrow 0$ limit can be taken without any rescaling, so that the
action is now given by \cite{deBoer:2021jej}

\begin{equation}
  \begin{aligned}
    S\left[x^{\mu}(\zeta),e(\zeta)\right]
    & =
  -  \frac{\tilde{m}}{2} \int d\zeta \left[e^{-1}\dot{x}^{m}\dot{x}^{m}    +e\right]\, .
  \end{aligned}
\end{equation}

The action is invariant under the global transformations

\begin{equation}
  \delta x^{\mu}
  =
  a^{m}\pi_{m}{}^{\mu}+\tfrac{1}{2}\sigma^{mn}\pi_{mn}{}^{\mu}
  +b^{m}\varpi^{(-1)}_{m0}+c\varpi^{(-1)}_{0}\,,
\end{equation}
so that the algebra of the symmetries is the one of the Carroll group  Eq.~\eqref{Carrol}.

\subsection{The $p$-brane limits}
\label{sec-Minkowskipbrane}

Starting from the Minkowski metric we now define a $p$-brane limit by  rescaling the first $p+1$ Cartesian coordinates $x^{A}$, $A=0,\cdots,p$ with
a parameter $\rho$ as $x^{A}\equiv\rho y^{A}$ obtaining

\begin{equation}
  ds^{2}
  =
  \rho^{2}\eta_{AB}dy^{A}dy^{B} -dx^{m}dx^{m}\,,
  \hspace{1cm}
  A=0,\cdots p\,,
  m=1,\cdots, d-(p+1)\,.
\end{equation}
Here $\eta_{AB}$ is the Minkowski metric in the $(p+1)$-dimensional subspace
described by the coordinates $y^{A}$.
The two singular metrics that occur in the $\rho\rightarrow 0$ and
$\rho\rightarrow \infty$ limits are

\begin{equation}
  \label{eq:hmunuMinkpbrane}
  h_{\mu\nu}dx^{\mu}dx^{\nu}
  =
 -dx^{m}dx^{m}\,,
  \hspace{1cm}
  k_{\mu\nu}dx^{\mu}dx^{\nu}
  =
  \eta_{AB}dy^{A}dy^{B}\,.
\end{equation}
The Killing vectors of Minkowski spacetime in Cartesian coordinates are given
in Eqs.~(\ref{eq:KillingvectorsMinkowskiCartesian}) and their Lie algebra is
given in Eqs.~(\ref{eq:LieAlgebraKillingvectorsMinkowskiCartesian}).
They admit an expansion in the parameter $\rho$, so that they can be identified with the Killing vectors in the general framework as

\begin{subequations}
  \begin{align}
   \varpi_{A}
    & =
      \frac{1}{\rho}\partial_{A}\,,
    \\
    & \nonumber \\
     \pi_{m}
    & =
      \partial_{m}\,,
    \\
    & \nonumber \\
  \pi_{AB}
    & =
      2\eta_{AB}{}^{D}{}_{C}y^{C}\partial_{B}
      =
      -\eta_{AC}y^{C}\partial_{B}+\eta_{BC}y^{C}\partial_{A}\,,
        \\
    & \nonumber \\
    \varpi_{mA}
    & =
      \rho\eta_{AB}y^{B}\partial_{m} +\frac{1}{\rho}x^{m}\partial_{A}\,,
          \\
    & \nonumber \\
\pi_{m n}
    & =
      2\eta_{mn}{}^{p}{}_{q}x^{q}\partial_{p}
      =
      x^{m}\partial_{n}-x^{n}\partial_{m}\,,
  \end{align}
\end{subequations}
The non-vanishing Lie brackets of these Killing vectors are

\begin{subequations}
  \begin{align}
   [\pi_{mn}, \pi_{pq}]
    & =
    -\delta_{mp}\pi_{nq}
    -\delta_{nq}\pi_{mp}
    +\delta_{mq}\pi_{np}
    +\delta_{np}\pi_{mq}\,,
    \\
    & \nonumber \\
   [\pi_{AB}, \pi_{CD}]
    & =
    +\eta_{AC}\pi_{BD}
    +\eta_{BD}\pi_{AC}
    -\eta_{AD}\pi_{BC}
    -\eta_{BC}\pi_{AD}\,,
    \\
    & \nonumber \\
   [\pi_{m}, \pi_{np}]
    & =
2\delta_{m[n}\pi_{p]}\,,
    \\
    & \nonumber \\
   [ \pi_{mn},\varpi_{p A}]
    & =
      -2\delta_{p[m}\varpi_{n]A}\,,
      \\
    & \nonumber \\
   [\pi_{m},\varpi_{n A}]
    & =
      \delta_{mn}\varpi_{A}\,,
      \\
    & \nonumber \\
   [\pi_{AB},\varpi_{C}]
    & =
      2\eta_{C[A}\varpi_{B]}\,,
      \\
    & \nonumber \\
   [\pi_{AB},\varpi_{mC}]
    & =
      2\eta_{C[A|}\varpi_{m|B]}\,,
      \\
    & \nonumber \\
    [\varpi_{A},\varpi_{m B}]
    & =
      \eta_{AB}\pi_{m}\,,
    \\
    & \nonumber \\
    [\varpi_{mA},\varpi_{nB}]
    & =
     \eta_{AB}\pi_{mn}-\delta_{mn}\pi_{AB}\,,
  \end{align}
\end{subequations}

\noindent
which fits the general pattern given in Eq.~(\ref{eq:Liealgebrapattern}).

The subalgebra generated by the $\pi$s is the product of the Lorentz algebra
in $(p+1)$ dimensions and the algebra of the Euclidean group in $d-(p+1)$
dimensions.  It leaves invariant the two singular metrics

\begin{equation}
  \pounds_{\pi}k_{\mu\nu}
  =
  \pounds_{\pi}h_{\mu\nu}
  =
  0\,.
\end{equation}

\noindent
Defining

\begin{subequations}
  \begin{align}
  \varpi^{(-1)}_{A}
  & \equiv
    \partial_{A}\,,
    \\
    & \nonumber \\
  \varpi^{(-1)}_{m A}
  & =
x^{m}\partial_{A}\,,
    \\
    & \nonumber \\
  \varpi^{(1)}_{m A}
  & =
\eta_{AB}y^{B}\partial_{m}\,,
  \end{align}
\end{subequations}

\noindent
we find, in all cases,

\begin{equation}
  \pounds_{\varpi^{(-1)}}h_{\mu\nu}
  =
  \pounds_{\varpi^{(1)}}k_{\mu\nu}
  =
  0\,,
\end{equation}

\noindent
and

\begin{equation}
  \pounds_{\varpi^{(1)}_{mA}}h_{\mu\nu}
  +
  \pounds_{\varpi^{(-1)}_{mA}}k_{\mu\nu}
  =
  0\,,
  \hspace{1cm}
  \pounds_{\varpi^{(-1)}_{A}}k_{\mu\nu}
  =
  0\,.
\end{equation}
All the Killing vectors $\varpi^{(-1)}_{A}$ belong to the class of those which
do not have a $\varpi^{(1)}$ component and generate symmetries of $g,h$ and $k$.
They will  therefore survive all the limits.

\subsubsection{$\rho\rightarrow \infty$ with auxiliary  fields}

The $\rho\rightarrow\infty$ limit of the $p$-brane action in Minkowski
spacetime takes the general form in
Eq.~(\ref{eq:roinftylimitnorescalingsaction}) with $h_{\mu\nu}$ given in
Eq.~(\ref{eq:hmunuMinkpbrane}).
This action is invariant under the global transformations

\begin{equation}
  \delta x^{\mu}
  =
  a^{m}\pi_{m}{}^{\mu}+\tfrac{1}{2}\sigma^{mn}\pi_{mn}{}^{\mu}
  +\tfrac{1}{2}\sigma^{AB}\pi_{AB}{}^{\mu}
  +b^{mA}\varpi^{(1)}_{mA}+c^{A}\varpi^{(-1)}_{A}\,.
\end{equation}

\noindent
The algebra of symmetries of this action is that generated by the
$\{\pi_{m},\pi_{mn},\pi_{AB}\}$ generators (those of the Lorentz group in
$(p+1)$ dimensions and of the Euclidean group in $d-(p+1)$ dimensions)
together with the $\varpi^{(-1)}_{A}$ and $\varpi^{(1)}_{mA}$ generators,
which have the following non-trivial Lie brackets:

\begin{subequations}
  \label{eq:pbraneGalileialgebra}
  \begin{align}
  [\pi_{mn},\varpi^{(1)}_{pA}]
  & =
  -2\delta_{p[m}\varpi^{(1)}_{n]A}\,,
    \\
    & \nonumber \\
  [\pi_{AB},\varpi^{(-1)}_{C}]
 & =
  +2\eta_{C[A}\varpi^{(-1)}_{B]}\,,
    \\
    & \nonumber \\
  [\pi_{AB},\varpi^{(1)}_{mC}]
 & =
  +2\eta_{C[A|}\varpi^{(1)}_{m|B]}\,,
    \\
    & \nonumber \\
  [\varpi^{(-1)}_{A},\varpi^{(1)}_{mB}]
  & =
    \eta_{AB}\pi_{m}\,,
  \end{align}
\end{subequations}
This is the algebra of the \textit{$p$-brane Galilei group}.

\subsubsection{$\rho\rightarrow 0$ with auxiliary fields}

As indicated in Section~\ref{sec-rhozerolimits}, the action can be obtained by
interchanging the degenerate metrics $h$ and $k$, see
Eq.~(\ref{eq:hmunuMinkpbrane}), in the action
Eq.~(\ref{eq:roinftylimitnorescalingsaction}), replacing simultaneously $\rho$
by $1/\rho$.
The resulting action is invariant under the global transformations

\begin{equation}
  \delta x^{\mu}
  =
  a^{m}\pi_{m}{}^{\mu}+\tfrac{1}{2}\sigma^{mn}\pi_{mn}{}^{\mu}
  +\tfrac{1}{2}\sigma^{AB}\pi_{AB}{}^{\mu}
  +b^{mA}\varpi^{(-1)}_{mA}+c^{A}\varpi^{(-1)}_{A}\,.
\end{equation}
The algebra of symmetries is that generated by the
$\{\pi_{m},\pi_{mn},\pi_{AB}\}$ generators (those of the Lorentz group in
$(p+1)$ dimensions and of the Euclidean group in $d-(p+1)$ dimensions)
together with the $\{\varpi^{(-1)}_{A},\varpi^{(-1)}_{mA}\}$ generators, which
only have the following non-trivial Lie brackets with the $\pi$s:

\begin{subequations}
  \label{eq:pbraneCarrollalgebra}
  \begin{align}
  [\pi_{mn},\varpi^{(-1)}_{pA}]
  & =
  -2\delta_{p[m}\varpi^{(-1)}_{n]A}\,,
    \\
    & \nonumber \\
  [\pi_{m},\varpi^{(-1)}_{nA}]
  & =
    \delta_{mn}\varpi^{(-1)}_{A}\,,
    \\
    & \nonumber \\
  [\pi_{AB},\varpi^{(-1)}_{C}]
 & =
  +2\eta_{C[A}\varpi^{(-1)}_{B]}\,,
    \\
    & \nonumber \\
  [\pi_{AB},\varpi^{(-1)}_{mC}]
 & =
  +2\eta_{C[A|}\varpi^{(-1)}_{m|B]}\,,
  \end{align}
\end{subequations}
This is the algebra of the \textit{$p$-brane Carroll group}. It is worth
comparing this algebra with the $p$-brane Galilei algebra in
Eq.~(\ref{eq:pbraneGalileialgebra}). As mentioned at the end of
Section~\ref{sec-rhozerolimits}, the subalgebra generated by the $\pi$s is
common to the two limiting algebras and the rest of the relevant commutators
are related by the $1\leftrightarrow -1$ duality (which acts on all indices
except on those of $\varpi^{(-1)}_{A}$, which survive in both limits). In
Table~\ref{figure1} we compare the Lie brackets of the $p$-brane Galilei and
Carroll algebras from the point of view of this $1\leftrightarrow -1$
duality. For $p=0$ a geometric interpretation of this duality has been
discussed in \cite{Duval:2014uoa}.

\begin{table}[h]
  \centering
  \begin{tabular}{|c|c|}
    \hline
    $p$-brane Galilei & $p$-brane Carroll \\
    \hline
    & \\
$[\pi_{mn},\varpi^{(1)}_{pA}]
  =
    -2\delta_{p[m}\varpi^{(1)}_{n]A}\,,$
            &
 $[\pi_{mn},\varpi^{(-1)}_{pA}]
  =
              -2\delta_{p[m}\varpi^{(-1)}_{n]A}\,,$
    \\
            & \\
    \hline
    & \\
    $[\pi_{AB},\varpi^{(-1)}_{C}]
  =
    2\eta_{C[A}\varpi^{(-1)}_{B]}\,,$
            &
    $[\pi_{AB},\varpi^{(-1)}_{C}]
  =
    2\eta_{C[A}\varpi^{(-1)}_{B]}\,,$
\\
            & \\
    \hline
    & \\
    $[\pi_{AB},\varpi^{(1)}_{mC}]
  =
    2\eta_{C[A|}\varpi^{(1)}_{m|B]}\,,$
            &
    $[\pi_{AB},\varpi^{(-1)}_{mC}]
  =
    2\eta_{C[A|}\varpi^{(-1)}_{m|B]}\,,$
\\
            & \\
    \hline
    & \\
    $[\varpi^{(-1)}_{A},\varpi^{(1)}_{mB}]
  =
    \eta_{AB}\pi_{m}\,,$
            &
        $[\varpi^{(-1)}_{A},\varpi^{(-1)}_{mB}]
  =
    0\,,$
\\
            & \\
    \hline
            & \\
    $[\pi_{m},\varpi^{(1)}_{nA}]
  =
  0\,,$
            &
  $[\pi_{m},\varpi^{(-1)}_{nA}]
  =
              \delta_{mn}\varpi^{(-1)}_{A}\,.$
    \\
            & \\
    \hline
  \end{tabular}
  \caption{\small This table compares the Lie brackets of the $p$-brane
    Galilei and $p$-brane Carroll algebras for the same $p$ from the
    $1\leftrightarrow -1$ duality point of view.  The brackets in the first
    line are related by the $1\leftrightarrow -1$ duality. The brackets in the
    second line are related by the same duality, which does not act on the
    $\varpi^{(-1)}_{A}$ generators, which are common to both algebras. Then,
    in the third line the right hand side of the Carroll bracket must vanish
    because it has two $-1$ indices. Something similar happens in the last
    line: if one starts with the Carroll bracket and applies the duality
    rule, then the dual Galilei bracket must vanish because it would have
    opposite $1,-1$ indices at both sides of the bracket relation.}
\label{figure1}
\end{table}

There exists a different formal duality between the $p$-brane Galilei algebra
and the $d-p-2$-brane Carroll algebra
\cite{Barducci:2018wuj,Bergshoeff:2020xhv}. The whole set of non-vanishing Lie
brackets of these two Lie algebras are re-arranged in Table~\ref{figure1other}
so as to make manifest the formal duality, sometimes called brane
$\leftrightarrow$ dual brane duality, that exists between these two algebras
under the interchange of the $A,B,\ldots$ and $m,n,\ldots$ indices. In this
case, in contrast to the $1\leftrightarrow -1$ mapping, the correspondence
between the generators of the algebra is not 1 to 1 unless we compare the
$p$-brane Galilei and the $p'=d-(p+2)$-brane Carroll algebras in which case
the range of the $A,B$ indices is $p^{\prime} +1=d-p-1$ and the range of the
$m,n$ is $d-p^{\prime} -1 =p +1$.

\begin{table}[h!]
  \centering
  \begin{tabular}{|c|c|}
    \hline
$p$-brane    Galilei & dual $(d-p-2)$-brane Carroll \\
    \hline
    & \\
         $[\pi_{mn}, \pi_{pq}]
    =
    +\eta_{mp}\pi_{nq}
    +\eta_{nq}\pi_{mp}\ldots
    $
            &
       $[\pi_{AB}, \pi_{CD}]
    =
    +\eta_{AC}\pi_{BD}
    +\eta_{BD}\pi_{AC}\ldots
              \,,$
    \\
                     & \\
    \hline
    & \\
       $[\pi_{AB}, \pi_{CD}]
    =
    +\eta_{AC}\pi_{BD}
    +\eta_{BD}\pi_{AC}\ldots
              \,,$
            &
         $[\pi_{mn}, \pi_{pq}]
    =
    +\eta_{mp}\pi_{nq}
    +\eta_{nq}\pi_{mp}\ldots
    $
    \\
            & \\
    \hline
    & \\
$[\pi_{mn},\varpi^{(1)}_{pA}]
  =
    2\eta_{p[m}\varpi^{(1)}_{n]A}\,,$
            &
$[\pi_{AB},\varpi^{(-1)}_{mC}]
  =
    2\eta_{C[A|}\varpi^{(-1)}_{m|B]}\,,$
    \\
            & \\
    \hline
    & \\
$[\pi_{AB},\varpi^{(1)}_{mC}]
  =
    2\eta_{C[A|}\varpi^{(1)}_{m|B]}\,,$
            &
 $[\pi_{mn},\varpi^{(-1)}_{pA}]
  =
              2\eta_{p[m}\varpi^{(-1)}_{n]A}\,,$
    \\
            & \\
    \hline
    & \\
    $[\pi_{AB},\varpi^{(-1)}_{C}]
  =
    2\eta_{C[A}\varpi^{(-1)}_{B]}\,,$
            &
   $[ \pi_{mn},\pi_{p},]
     =
2\eta_{p[m}\pi_{n]}\,,$
    \\
            & \\
    \hline
    & \\
   $[ \pi_{mn},\pi_{p},]
     =
2\eta_{p[m}\pi_{n]}\,,$
            &
    $[\pi_{AB},\varpi^{(-1)}_{C}]
  =
    2\eta_{C[A}\varpi^{(-1)}_{B]}\,,$
    \\
            & \\
    \hline
            & \\
    $[\pi_{m},\varpi^{(1)}_{nA}]
  =
  0\,,$
            &
        $[\varpi^{(-1)}_{A},\varpi^{(-1)}_{mB}]
  =
    0\,,$
    \\
            & \\
    \hline
    & \\
    $[\varpi^{(-1)}_{A},\varpi^{(1)}_{mB}]
  =
    \eta_{AB}\pi_{m}\,,$
            &
  $[\pi_{m},\varpi^{(-1)}_{nA}]
  =
              -\eta_{mn}\varpi^{(-1)}_{A}\,.$
\\
            & \\
    \hline
  \end{tabular}
  \caption{\small This table compares the $p$-brane Galilei and Carroll algebras from a different so-called brane $\leftrightarrow$ dual brane point of view. This is a formal duality which relates the different generators by interchanging
    the  $A,B,\ldots$ and $m,n,\ldots$ indices. This duality is one to one provided one compares a $p$-brane Galilei algebra with a dual $d-p-2$-brane Carroll algebra. }
\label{figure1other}
\end{table}

\subsubsection{$\rho\rightarrow \infty$ with rescalings}


We can remove the divergence in the action by performing the  rescaling

\begin{equation}
  \gamma_{ij}\rightarrow c^{2}\gamma_{ij}\,,
  \hspace{1cm}
  T\rightarrow c^{-(p+1)}T\,,
\end{equation}
and then taking the limit. The action involves the $(p+1)$-dimensional subspace with coordinates $y^{A}$ and it turns to be invariant under

\begin{equation}
  \delta x^{\mu}
  =
  a^{m}\pi_{m}{}^{\mu}+\tfrac{1}{2}\sigma^{mn}\pi_{mn}{}^{\mu}
  +\tfrac{1}{2}\sigma^{AB}\pi_{AB}{}^{\mu}
  +b^{mA}\varpi^{(1)}_{mA}+c^{A}\varpi^{(-1)}_{A}\, ,
\end{equation}
which corresponds to the $p$-brane Galilei group  Eq.~\eqref{eq:pbraneGalileialgebra}.

\subsubsection{$\rho\rightarrow 0$ without rescalings}

The $\rho\rightarrow 0$ limit can be taken without any rescaling and the
action involves the $d-(p+1)$-dimensional subspace described by the
coordinates $x^m$. The action is now invariant under the $p$-brane Carroll
symmetry Eq.~\eqref{eq:pbraneCarrollalgebra}.

\section{Example 2: Limits of Anti-de Sitter spacetime}
\label{sec-holographiclimit}

In this section we are going to apply our general techniques to Anti-de Sitter
(AdS) spacetime and explore several new and unconventional singular limits. We
will first discuss the so-called holographic limits in which the Lorentz group
of the holographic screen\footnote{Since the holographic screen has
  codimension 1, the Lorentz group is not the one associated to the dimension
  of the original spacetime, but a smaller one.} together with dilatations
survive as isometry algebra\footnote{As usual, all the symmetries of the
  relativistic particle and $p$-brane actions ``survive'' in the sense that
  they are in one-to-one correspondence with a symmetry of the actions that
  arise after the limits.} and, after that, examples of $p$-brane limits.

\subsection{Holographic Limits}

Our starting point is the metric of AdS$_{p+2}$ in horospheric
coordinates\footnote{Some technical details about these coordinates can be
  found in Appendix~\ref{sec-ADS}, around
  Eqs.~(\ref{eq:horosphericcoordinates}).}. We use unhatted Greek indices
$\mu,\nu,\ldots=0,1,\cdots,p$ for the Cartesian coordinates $x^{\mu}$ of the
holographic screen and the holographic $(p+1)$\textsuperscript{th} coordinate
$x^{p+1}\equiv z$ normal to it.  We use hatted Greek indices
$\hat{\mu},\hat{\nu},\dots = 0,\cdots,p+1$ to indicate the horospheric
coordinates $x^{\hat{\mu}}=(x^{\mu},z)$. Thus, the AdS$_{p+2}$ spacetime is
described by the metric

\begin{equation}
    d\hat{s}^{2}_{p+2}
    =
    \left(\frac{R}{z}\right)^{2}
    \left[\eta_{\mu\nu}dx^{\mu}dx^{\nu}-dz^{2}\right]\, ,
\end{equation}

\noindent
which is invariant under the group SO$(2,p+1)$, whose generators are labeled
by an antisymmetric pair of double-hatted Greek indices
$\hat{\hat{\alpha}},\hat{\hat{\beta}},\ldots=+,-,\mu$. Hence, there are four
kinds of Killing vectors
$\hat{\hat{k}}_{\hat{\hat{\alpha}}\hat{\hat{\beta}}}$, which are

\begin{itemize}
\item The generators of the Lorentz group of the Minkowski metric associated
  to the holographic screen.

\item $\hat{\hat{k}}_{+\alpha}$, which generate
  translations in the same metric.
  
\item $\hat{\hat{k}}_{-\alpha}$, which generate
  conformal transformations of that metric compensated by rescalings of the
  holographic coordinate $z$.

\item $\hat{\hat{k}}_{-+}$ which generate dilatations of all the
  coordinates\footnote{See Appendix~\ref{sec-ADS} for the explicit form of the
    Killing vectors}.
\end{itemize}

We are interested in taking limits in which the radius $R$, which here will
play the role of the generic parameter $\rho$, goes to
$\left(0,\infty\right)$. However, in order to apply our machinery, we first
need to bring the metric in the desired form. This can be achieved by
rescaling the holographic coordinate $z$ defining a new rescaled coordinate
$w$ by 

\begin{equation}
z=Rw\,.
\end{equation}

After this rescaling, the metric takes form

\begin{equation}
  \label{eq:ADShorospheric2}
    d\hat{s}^{2}_{p+2}
    =
    \frac{1}{w^{2}}
    \eta_{\mu\nu}dx^{\mu}dx^{\nu}-R^{2}\frac{dw^{2}}{w^{2}}\,.
\end{equation}

\noindent
We can rewrite this metric in the form of Eq.~(\ref{eq:assumptionmetric}). In
this case, the two singular metrics, $\hat{h}_{\hat{\mu}\hat{\nu}}$ and
$ \hat{k}_{\hat{\mu}\hat{\nu}}$ have ranks $p+1$ and $1$ and signatures
$(+,-,\cdots,-,0)$ and $(0,\cdots,0,-)$, respectively, with the AdS radius
parameter $R$ playing the role of the generic parameter $\rho$:

\begin{equation}
  \label{eq:decompositionmetric}
  d\hat{s}^{2}_{p+2}
  =
  \hat{h}_{\hat{\mu}\hat{\nu}}d\hat{x}^{\hat{\mu}}d\hat{x}^{\hat{\nu}}
  +R^{2}\hat{k}_{\hat{\mu}\hat{\nu}}d\hat{x}^{\hat{\mu}}d\hat{x}^{\hat{\nu}}\,,
\end{equation}

\noindent
with

\begin{subequations}
  \begin{align}
    \hat{h}_{\hat{\mu}\hat{\nu}}d\hat{x}^{\hat{\mu}}d\hat{x}^{\hat{\nu}}
    & =
    \frac{1}{w^{2}}
      \eta_{\mu\nu}dx^{\mu}dx^{\nu}\,,
    \\
    & \nonumber \\
    \hat{k}_{\hat{\mu}\hat{\nu}}d\hat{x}^{\hat{\mu}}d\hat{x}^{\hat{\nu}}
    & =
    -\frac{dw^{2}}{w^{2}}\,.
  \end{align}
\end{subequations}

\noindent
Notice that, since the metric $\hat{h}_{\mu\nu}$ is proportional to the
$(p+1)$-dimensional Minkowski metric, it will be invariant under the
$(p+1)$-dimensional Lorentz group SO$(1,p)$ (actually, under the whole
Poincar\'e group in $(p+1)$ dimensions). The metric is Lorentzian or
relativistic in this restricted sense, but non-Lorentzian or non-relativistic
when all the space and time dimensions are taken into account.

In this setup, in the $R\rightarrow 0$ for a constant value of the coordinate
$w$, we will be getting close to the origin ($z\rightarrow 0$) while the
curvature of the AdS spacetime $\sim 1/R^{2}$ grows and the paraboloid
degenerates into a sort of cone. In the $R\rightarrow\infty$ limit, for a
constant value of the coordinate $w$, we will be getting close to the boundary
$z\rightarrow \infty$ while the curvature decreases because the paraboloid
grows in size.

In these new coordinates, the Killing vectors take the form

\begin{subequations}
  \label{eq:ADSKillingvectorshorospheric2}
  \begin{align}
    \hat{\hat{\xi}}_{\alpha\beta}
    & =
      2\eta_{\alpha\beta}{}^{\mu}{}_{\nu} x^{\nu}\partial_{\mu}\,,
    \\
    & \nonumber \\
    \hat{\hat{\xi}}_{+\alpha}
    & =
      \frac{R}{2}\partial_{\alpha}\,,
    \\
    & \nonumber \\
    \hat{\hat{\xi}}_{-\alpha}
    & =
      \frac{1}{2R} \left\{
      x\cdot x\partial_{\alpha}
      -2\eta_{\alpha\mu}x^{\mu}(x^{\nu}\partial_{\nu}+w\partial_{w})
      \right\}
      - \frac{R}{2} w^{2}\partial_{\alpha} \,,
    \\
    & \nonumber \\
    \hat{\hat{\xi}}_{-+}
    & =
    \tfrac{1}{2}\left(  x^{\mu}\partial_{\mu}+w\partial_{w}\right)\,.
  \end{align}
\end{subequations}

\noindent
Following the general framework explained in
Section~\ref{sec-Liealgebraofisometries}, we relabel these Killing vectors,
according to their dependence on the parameter $R$, as follows:

\begin{subequations}
  \label{eq:ADSKillingvectorshorospheric3}
  \begin{align}
    \pi_{\alpha\beta}
    & =
      2\eta_{\alpha\beta}{}^{\mu}{}_{\nu} x^{\nu}\partial_{\mu}\,,
    \\
    & \nonumber \\
    \pi
    & =
      \tfrac{1}{2}\left(  x^{\mu}\partial_{\mu}+w\partial_{w}\right)\,,
    \\
    & \nonumber \\
    \varpi_{+\alpha}
    & =
      \frac{R}{2}\partial_{\alpha}\,,
    \\
    & \nonumber \\
    \varpi_{-\alpha}
    & =
      \frac{1}{2R} \left\{
      x\cdot x\partial_{\alpha}
      -2\eta_{\alpha\mu}x^{\mu}(x^{\nu}\partial_{\nu}+w\partial_{w})
      \right\}
      - \frac{R}{2} w^{2}\partial_{\alpha}\,.
  \end{align}
\end{subequations}

The Lie brackets of these Killing vectors  can be read from
Eqs.~(\ref{eq:ADSalgebrahorospheric}). They  take  the form

\begin{subequations}
  \label{eq:ADSalgebrahorospheric2}
  \begin{align}
  [\pi_{\mu\nu},\pi_{\rho\sigma}]
  & =
  \eta_{\mu\rho}\pi_{\nu\sigma}
  +\eta_{\nu\sigma}\pi_{\mu\rho}
  -\eta_{\mu\sigma}\pi_{\nu\rho}
  -\eta_{\nu\rho}\pi_{\mu\sigma}\,,
    \\
    & \nonumber \\
  [\pi_{\mu\nu},\varpi_{\pm\alpha}]
  & =
    2\eta_{\alpha[\mu|}\varpi_{\pm|\nu]}\,,
    \\
    & \nonumber \\
  [\pi,\varpi_{\pm\alpha}]
  & =
  \pm\tfrac{1}{2}\varpi_{\pm\alpha}\,,
    \\
    & \nonumber \\
  [\varpi_{+\alpha},\varpi_{-\beta}]
  & =
  \tfrac{1}{2}\pi_{\alpha\beta}
  -\eta_{\alpha\beta}\pi\,,
  \end{align}
\end{subequations}

\noindent
which conforms to the general pattern given in
Eq.~(\ref{eq:Liealgebrapattern}).

We can now consider several different limits. For simplicity, we restrict to
the sigma model action of a massive point particle moving in the AdS$_{p+2}$
spacetime. The transformations that leave invariant such an action are

\begin{equation}
  \label{eq:transformations}
  \begin{aligned}
    \delta x^{\mu}
    & =
    \tfrac{1}{2}\sigma^{\alpha\beta}\pi_{\alpha\beta}{}^{\mu}
+2a^{\alpha}\varpi_{+\alpha}{}^{\mu}
+2b^{\alpha}\varpi_{-\alpha}{}^{\mu}
+2c\pi^{\mu}
\\
& \\
& =
\sigma^{\mu}{}_{\nu}x^{\nu} +R a^{\mu}
+\frac{1}{R}\left(x\cdot x b^{\mu} -2 b\cdot x x^{\mu}\right)
-Rw^{2}b^{\mu} +cx^{\mu}\,,
\\
& \\
\delta w
& =
2b^{\alpha}\varpi_{-\alpha}{}^{w}
+c\pi^{w}=
-\frac{2}{R} b\cdot x w+cw\,.
  \end{aligned}
\end{equation}

We need to rescale the parameters $\sigma^{\mu\nu},a^{\mu},b^{\mu},c$ to keep
these transformation rules finite in the $R\rightarrow \left(0,\infty\right)$
limits. We can always preserve the symmetries corresponding to the generators
$\varpi_{+\alpha}$ by absorbing $R$ into $a^{\mu}$ since these generators
expand only in a term proportional to $R$:

\begin{equation}
  \varpi_{+\alpha} = R \varpi^{(1)}_{+\alpha}\,,
  \hspace{1cm}
  \varpi^{(1)}_{+\alpha}
  =
  \tfrac{1}{2}\partial_{\alpha}\,.
\end{equation}

\noindent
All the symmetries corresponding to the $\varpi_{-\alpha}$s, which can be
written as 

\begin{equation}
  \begin{aligned}
    \varpi_{-\alpha}
    & =
    \frac{1}{R}\varpi^{(-1)}_{-\alpha}
    +R\varpi^{(1)}_{-\alpha}\,,
    \\
    & \\
    \varpi^{(-1)}_{-\alpha}
    & =
    \tfrac{1}{2} \left\{ x\cdot x\partial_{\alpha}
      -2\eta_{\alpha\mu}x^{\mu}(x^{\nu}\partial_{\nu}+w\partial_{w})
    \right\}\,,
    \\
    & \\
    \varpi^{(1)}_{-\alpha}
    & =
    -\tfrac{1}{2} w^{2}\partial_{\alpha}\,,
  \end{aligned}
\end{equation} 

\noindent
will also be preserved but they will be generated by either
$ \varpi^{(1)}_{-\alpha}$ or $ \varpi^{(-1)}_{-\alpha}$ which are Killing
vectors of the metrics $k_{\mu\nu}$ and $h_{\mu\nu}$ respectively and not of
the the full AdS$_{p+2}$ metric.

We will now discuss the different limits.

\subsubsection{$R\rightarrow\infty$ with auxiliary fields}

Following the general procedure, we obtain the following action including an
auxiliary field $\lambda$:

\begin{equation}
  \label{eq:ADShorospheric2actionlambda1limit}
    S\left[x^{\mu}(\zeta),w(\zeta)\right]
     =
    -\frac{m}{2} \int d\zeta e\left\{e^{-2}
      \left[ \frac{ \eta_{\mu\nu}\dot{x}^{\mu}\dot{x}^{\nu}}{w^{2}}
        +\lambda\frac{\dot{w}}{w}\right]
      + 1\right\}\,.
\end{equation}

\noindent
This action is invariant under the following transformations:

\begin{equation}
  \label{eq:Rinftytransformations}
  \begin{aligned}
    \delta x^{\mu}
    & =
\sigma^{\mu}{}_{\nu}x^{\nu} +a^{\mu}
-w^{2}b^{\mu} +cx^{\mu}\,,
\\
& \\
\delta w
& =
cw\,,
\\
& \\
\delta\lambda
& =
 4 b\cdot \dot{x}\,.
  \end{aligned}
\end{equation}

The generators of these symmetries are given by
$\{\pi_{\alpha\beta},\pi,\omega_{+\alpha}^{(1)},\omega_{-\alpha}^{(1)}\}$. The
$\pi$ generators (the only Killing vectors of the three metrics, the original
and the two singular ones) correspond to the product of the
$(p+1)$-dimensional Lorentz algebra with dilatations. The non-vanishing Lie
brackets involving the $\varpi$ generators are given by

\begin{equation}
      \label{eq:Rinfinityalgebra}
  [\pi_{\alpha\beta},\omega_{+\gamma}^{(1)}]
  =
  2\eta_{\gamma[\alpha|}\omega_{+|\beta]}^{(1)}\,,
  \hspace{.7cm}
  [\pi_{\alpha\beta},\omega_{-\alpha}^{(1)}]
  =
  2\eta_{\gamma[\alpha|}\omega_{-|\beta]}^{(1)}\,,
  \hspace{.7cm}
  [\pi,\omega_{\pm\alpha}^{(1)}]
  =
  \mp\tfrac{1}{2}\omega_{\pm\alpha}^{(1)}\,,
\end{equation}

\noindent
and, therefore, the $\omega_{+\alpha}^{(1)},\omega_{-\alpha}^{(1)}$ generators
can be interpreted as two momenta with opposite weights under dilatations
which commute with each other.  As usual, the number of generators of the
symmetry algebra is the dimensional of the isometry group of the original
metric AdS$_{p+2}$, $SO(2,p+1)$, namely $(p+3)(p+2)/2$.

\subsubsection{$R\rightarrow 0$ with auxiliary fields}

If we use the rescaled action Eq.~(\ref{eq:ADShorospheric2action2}), we have a
divergent term in the $R\rightarrow 0$ limit. We can deal with this problem by
performing a Hubbard-Stratonovich transformation introducing an auxiliary
variable $\lambda$. We thus obtain

\begin{equation}
  \label{eq:ADShorospheric2action2lambda}
    S\left[x^{\mu}(\zeta),w(\zeta),\lambda(\zeta)\right]
     =
    -\frac{m}{2} \int d\zeta e\left\{e^{-2}
      \left[ -\frac{R^{2}}{4}\lambda^{2}
        -\lambda\frac{\sqrt{\dot{x}\cdot\dot{x}}}{w}
        -\left(\frac{\dot{w}}{w}\right)^{2}\right] + 1\right\}\,.
\end{equation}
The equation of the auxiliary field $\lambda$ is solved by

\begin{equation}
  \lambda
  =
  -\frac{2}{R^{2}}\frac{\sqrt{\dot{x}\cdot\dot{x}}}{w}\,.
\end{equation}

\noindent
The field $\lambda$ must transform as the right-hand side of this equation if the
action is going to enjoy the same symmetries as the original one. Thus,

\begin{equation}
  \begin{aligned}
    \delta \lambda
    & =
    -\frac{2}{R^{2}}\delta \frac{\sqrt{\dot{x}\cdot\dot{x}}}{w}
    =
    \frac{2}{R}
    \frac{b\cdot\dot{x}\dot{w}}{\sqrt{\dot{x}\cdot\dot{x}}}\,.
  \end{aligned}
\end{equation}

If we now we take the $R\rightarrow 0$ limit we obtain the following action

\begin{equation}
  \label{eq:ADShorospheric2action2lambdalimit}
    S\left[x^{\mu}(\zeta),w(\zeta),\lambda(\zeta)\right]
     =
    \frac{m}{2} \int d\zeta e\left\{e^{-2}
      \left[
        \lambda\frac{\sqrt{\dot{x}\cdot\dot{x}}}{w}
        +\left(\frac{\dot{w}}{w}\right)^{2}\right] - 1\right\}\,,
\end{equation}

\noindent
and  transformation rules

\begin{equation}
  \label{eq:R0transformations}
  \begin{aligned}
    \delta x^{\mu}
    & =
\sigma^{\mu}{}_{\nu}x^{\nu} +a^{\mu}
+\left(x\cdot x b^{\mu} -2 b\cdot x x^{\mu}\right)
+\frac{c}{2}x^{\mu}\,,
\\
& \\
\delta w
& =
-2b\cdot x w+\frac{c}{2}w\,,
\\
& \\
\delta \lambda
& =
    4 \frac{b\cdot\dot{x}\dot{w}}{\sqrt{\dot{x}\cdot\dot{x}}}\,.
  \end{aligned}
\end{equation}

Thus, the action is invariant under so$(d-1)$ transformations, with generators
\,\,
$\{\pi_{\alpha\beta},\pi,\omega_{+\alpha}^{(1)},\omega_{-\alpha}^{(-1)}\}$
whose non-vanishing Lie brackets are given by

\begin{subequations}
  \begin{align}
    [\pi_{\alpha\beta},\omega_{+\gamma}^{(1)}]
    &=
      2\eta_{\gamma[\alpha|}\omega_{+|\beta]}^{(1)}\,,
    \\
    & \nonumber \\
    [\pi_{\alpha\beta},\omega_{-\alpha}^{(-1)}]
    &=
      2\eta_{\gamma[\alpha|}\omega_{-|\beta]}^{(-1)}\,,
    \\
    & \nonumber \\
    [\pi,\omega_{\pm\alpha}^{(\pm1)}]
    &=\mp
      \tfrac{1}{2}\omega_{\pm\alpha}^{(\pm1)}\,,
    \\
    & \nonumber \\
    [\varpi_{+\alpha}^{(1)},\varpi_{-\beta}^{(-1)}]
    &=
      \tfrac{1}{2}\pi_{\alpha\beta}
      -\eta_{\alpha\beta}\pi\,.
  \end{align}
\end{subequations}

This is nothing but the AdS$_{p+2}$ algebra so$(2,p+1)$ written as in
Eq.~(\ref{eq:ADSalgebrahorospheric2}), which is fully preserved in this limit.

The invariance of the action Eq.~(\ref{eq:ADShorospheric2actionlimit}) under
these transformations is manifest, except for the symmetries with parameters
$b^{\mu}$. Using the fact that under these symmetries we have

\begin{equation}
  \begin{aligned}
    \delta \frac{\sqrt{\dot{x}\cdot\dot{x}}}{w}
    =
    0\,,
  \end{aligned}
\end{equation}

\noindent
one can easily show that the transformation of $\lambda$, which has become a
Lagrange multiplier, is such that the new action obtained after taking the
limit is invariant.

\subsubsection{$R\rightarrow\infty$ with rescalings}

In order to take the  $R\rightarrow \infty$ limit of the action

\begin{equation}
  \label{eq:ADShorospheric2action}
    S\left[x^{\mu}(\zeta),w(\zeta)\right]
     =
    -\frac{m}{2} \int d\zeta e\left\{e^{-2}
      \left[ \frac{\eta_{\mu\nu}\dot{x}^{\mu}\dot{x}^{\nu}}{w^{2}}
        -R^{2}\left(\frac{\dot{w}}{w}\right)^{2}\right] + 1\right\}\,.
\end{equation}

\noindent
we make the following rescalings

\begin{equation}
  e\rightarrow R e\,,
  \hspace{1cm}
  m\rightarrow \frac{m}{R}\,,
\end{equation}

\noindent
such that the action takes the form

\begin{equation}
  \label{eq:ADShorospheric2action2}
    S\left[x^{\mu}(\zeta),w(\zeta)\right]
     =
    -\frac{m}{2} \int d\zeta e\left\{e^{-2}
      \left[ \frac{1}{R^{2}}\frac{\eta_{\mu\nu}\dot{x}^{\mu}\dot{x}^{\nu}}{w^{2}}
        -\left(\frac{\dot{w}}{w}\right)^{2}\right] + 1\right\}\,.
\end{equation}
Now taking the  $R\rightarrow \infty$ limit leaves us with

\begin{equation}
  \label{eq:ADShorospheric2action2limit}
    S\left[w(\zeta)\right]
     =
    \frac{m}{2} \int d\zeta e\left\{e^{-2}
      \left(\frac{\dot{w}}{w}\right)^{2}- 1\right\}\,,
\end{equation}

\noindent
which should be compared to eq.~(\ref{eq:ADShorospheric2actionlambda1limit}).

The action  Eq.~\eqref{eq:ADShorospheric2action2limit} is invariant under the
transformations given in eqs.(\ref{eq:Rinftytransformations}). The invariance
under many of those transformations is due to the elimination of the
$x^{\mu}(\zeta)$ coordinates which makes these symmetries trivial.

\subsubsection{$R\rightarrow 0$ without rescalings}

The $R\rightarrow 0$ limit can be taken directly in
Eq.~(\ref{eq:ADShorospheric2action}) and gives

\begin{equation}
  \label{eq:ADShorospheric2actionlimit}
    S\left[x^{\mu}(\zeta),w(\zeta)\right]
     =
    -\frac{m}{2} \int d\zeta e\left\{e^{-2}
      \frac{\eta_{\mu\nu}\dot{x}^{\mu}\dot{x}^{\nu}}{w^{2}}
         + 1\right\}\,.
\end{equation}
This action is invariant under the transformations

\begin{equation}
  \label{eq:R0transformationsnolambda}
  \begin{aligned}
    \delta x^{\mu}
    & =
\sigma^{\mu}{}_{\nu}x^{\nu} +a^{\mu}
+\left(x\cdot x b^{\mu} -2 b\cdot x x^{\mu}\right)
+cx^{\mu}\,,
\\
& \\
\delta w
& =
-2b\cdot x w+cw\,.
  \end{aligned}
\end{equation}

The generators of these symmetries are given by
$\{\pi_{\alpha\beta},\pi,\omega_{+\alpha}^{(1)},\omega_{-\alpha}^{(-1)}\}$. The
$\pi$ generators correspond to  the product of the $(p+1)$-dimensional Lorentz algebra with
dilatations. To compare with the $R\to\infty$  limit result given in eq.~(\ref{eq:Rinfinityalgebra}), we give the non-vanishing Lie brackets involving the $\varpi$ generators
in the right column of Table~\ref{tab:fig2}.

\begin{table}[h]
  \centering
  \begin{tabular}{|c|c|}
    \hline
    $R\rightarrow \infty$  &  $R\rightarrow 0$ \\
    \hline
    & \\
   $[\pi_{\alpha\beta},\omega_{+\gamma}^{(1)}]
  =
  2\eta_{\gamma[\alpha|}\omega_{+|\beta]}^{(1)}\,,$
    &
      $[\pi_{\alpha\beta},\omega_{+\gamma}^{(1)}]
     =
    2\eta_{\gamma[\alpha|}\omega_{+|\beta]}^{(1)}\,,$
    \\
    & \\
    \hline
    & \\
    $[\pi_{\alpha\beta},\omega_{-\alpha}^{(1)}]
  =
  2\eta_{\gamma[\alpha|}\omega_{-|\beta]}^{(1)}\,,$
    &
      $[\pi_{\alpha\beta},\omega_{-\alpha}^{(-1)}]
     =
      2\eta_{\gamma[\alpha|}\omega_{-|\beta]}^{(-1)}\,,$
    \\
    & \\
    \hline
    & \\
    $[\pi,\omega_{+\alpha}^{(1)}]
  =
  -\tfrac{1}{2}\omega_{+\alpha}^{(1)}$
    &
      $[\pi,\omega_{+\alpha}^{(1)}]
     =
      -\tfrac{1}{2}\omega_{+\alpha}^{(1)}\,,$
    \\
    & \\
    \hline
    & \\
   $[\pi,\omega_{-\alpha}^{(1)}]
  =
    +\tfrac{1}{2}\omega_{-\alpha}^{(1)}\,,$
    &
      $[\pi,\omega_{-\alpha}^{(-1)}]
    =
      \tfrac{1}{2}\omega_{-\alpha}^{(-1)}\,,$
    \\
    & \\
    \hline
    & \\
    $[\omega_{+\alpha}^{(1)},\omega_{-\beta}^{(1)}]
     =
0\,,$
    &
      $[\omega_{+\alpha}^{(1)},\omega_{-\beta}^{(-1)}]
     =
      \tfrac{1}{2}\pi_{\alpha\beta} -\eta_{\alpha\beta}\pi\,,$
    \\
    & \\
    \hline
  \end{tabular}
  \caption{\small In the left column we give the non-trivial Lie brackets of
    the $\varpi$ generators obtained in the $R\rightarrow \infty$ limit,
    see Eq.~(\ref{eq:Rinfinityalgebra}). They are related by the
    $1\leftrightarrow -1$ duality to the Lie brackets obtained in the $R\to 0$ limit without rescalings.
    The duality does not apply
    to the $\omega_{+\alpha}^{(1)}$ generators, which are present in both
    algebras. The last bracket in the left column vanishes after the duality  interchange because there
    would be two $1$ indices in the left hand side. That is why there is no
    $[\varpi,\varpi]$ bracket in Eq.~(\ref{eq:Rinfinityalgebra}).}
  \label{tab:fig2}
\end{table}

\subsubsection{$R\rightarrow\infty$ plus WZ term}

We finally discuss one more way of taking the $R\to\infty$ limit that corresponds to option 3  discussed in the introduction. We will cancel the divergence by coupling the particle
 moving in AdS$_{p+1}$ to  a 1-form field via a Wess-Zumino (WZ) term. First, we switch the sign of the ``cosmological constant'' term
in the action so that it effectively describes a tachyon. Then we rescale
$e\rightarrow R e$ and $m\rightarrow Rm$ and obtain

\begin{equation}
  \label{eq:ADShorospheric2action3}
    S\left[x^{\mu}(\zeta),w(\zeta)\right]
     =
    -\frac{m}{2} \int d\zeta e\left\{e^{-2}
      \left[ \frac{\eta_{\mu\nu}\dot{x}^{\mu}\dot{x}^{\nu}}{w^{2}}
        -R^{2}\left(\frac{\dot{w}}{w}\right)^{2}\right] -R^{2}\right\}\,.
\end{equation}

\noindent
Now we add the following pullback of 1-form $A_{\hat{\mu}}\dot{x}^{\hat{\mu}}$ over the worldline, i.e.~a WZ term, with a  judiciously chosen divergent $R^{2}$ term such that all $R^{2}$ terms in the action form a complete square (see below):

\begin{equation}
  A_{\hat{\mu}}\dot{x}^{\hat{\mu}}
  =
  -\frac{m}{2}\left(\pm 2 R^{2} \frac{\dot{w}}{w} +m_{\hat{\mu}}\dot{x}^{\hat{\mu}}\right)\,.
\end{equation}

\noindent
The 1-form field $m_{\hat{\mu}}(\hat{x})$ must transform under the
isometries of AdS$_{p+1}$ according to

\begin{equation}
  \begin{aligned}
    \delta m_{\mu}
    & =
    -m_{\nu}\sigma^{\nu}{}_{\mu}-\frac{c}{2}m_{\mu}
    +\frac{1}{R}\left( m_{\mu}b\cdot x +m_{w}b_{\mu} w
    -2m_{\nu}b^{[\nu}x^{\rho]}\eta_{\rho\mu}\right)
          \pm 2 R  b_{\mu}\,,
\\
& \\
\delta m_{w}
& =
-\frac{c}{2} m_{w}
+Rm_{\nu}b^{\nu} w +\frac{1}{R} m_{w}b\cdot x\,,
  \end{aligned}
\end{equation}

\noindent
making the WZ term invariant. In this way, the addition of the new WZ term
preserves the symmetries of the original action.

The modified action whose limit we want to take can now be written as

\begin{equation}
  \label{eq:ADShorospheric2action3+WZ}
    S\left[x^{\mu}(\zeta),w(\zeta)\right]
     =
    -\frac{m}{2} \int d\zeta\left\{  e^{-1}
      \left[ \frac{\eta_{\mu\nu}\dot{x}^{\mu}\dot{x}^{\nu}}{w^{2}}
        -R^{2}\left(\frac{\dot{w}}{w}\mp e\right)^{2}\right]
      +m_{\hat{\mu}}\dot{x}^{\hat{\mu}}\right\}\,.
\end{equation}
Now we perform a Hubbard-Stratonovich transformation introducing an  auxiliary variable $\lambda$ thereby
rewriting the action in the equivalent form

\begin{equation}
  \label{eq:ADShorospheric2action3+WZlambda}
    S\left[x^{\mu}(\zeta),w(\zeta)\right]
     =
    -\frac{m}{2} \int d\zeta \left\{  e^{-1}
      \left[ \frac{\eta_{\mu\nu}\dot{x}^{\mu}\dot{x}^{\nu}}{w^{2}}
        +\frac{1}{4R^{2}}\lambda^{2}
        +\lambda\left(\frac{\dot{w}}{w}\mp e\right)\right]
      +m_{\hat{\mu}}\dot{x}^{\hat{\mu}}\right\}\,.
\end{equation}
The auxiliary field $\lambda$ must transform as the solution of its equation in order to
preserve the symmetries of the original action, that is

\begin{equation}
    \delta \lambda
    =
    -2R^{2}\delta \frac{\dot{w}}{w}
    =
    2R b\cdot\dot{x}\,.
\end{equation}

We next take the $R\rightarrow \infty$ in the action and in the transformations
with the rescalings of the transformation parameters used before. This leads
to the following action

\begin{equation}
  \label{eq:ADShorospheric2action3+WZlambdalimit}
    S\left[x^{\mu}(\zeta),w(\zeta)\right]
     =
    -\frac{m}{2} \int d\zeta\left\{ e^{-1}
      \left[ \frac{\eta_{\mu\nu}\dot{x}^{\mu}\dot{x}^{\nu}}{w^{2}}
        +\lambda\frac{\dot{w}}{w}\right]
      \mp \lambda +m_{\hat{\mu}}\dot{x}^{\hat{\mu}}\right\}\,,
\end{equation}

\noindent
and to the transformations Eqs.~(\ref{eq:Rinftytransformations}) plus

\begin{equation}
  \begin{aligned}
    \delta m_{\mu}
    & =
    -m_{\nu}\sigma^{\nu}{}_{\mu}-\frac{c}{2}m_{\mu}
          \pm 4 b_{\mu}\,,
\\
& \\
\delta m_{w}
& =
-\frac{c}{2} m_{w}+2m_{\nu}b^{\nu} w\,.
  \end{aligned}
\end{equation}

The action Eq.~(\ref{eq:ADShorospheric2action3+WZlambdalimit}) is the action
Eq.~(\ref{eq:ADShorospheric2actionlambda1limit}), which is invariant under all
the transformations, plus a term involving  the 1-form
$ \mp \lambda +m_{\hat{\mu}}\dot{x}^{\hat{\mu}}$  which also must be
invariant under all the transformations. This implies that

\begin{equation}
  \delta \lambda = \pm \delta (m_{\hat{\mu}}\dot{x}^{\hat{\mu}})\,.
\end{equation}

The equation for $\lambda$ is solved by

\begin{equation}
  e=\pm \frac{\dot{w}}{w}\,,
\end{equation}

\noindent
which we can substitute back in the action which now takes the form

\begin{equation}
  \label{eq:ADShorospheric2action3+WZlambdalimitnolambda}
    S\left[x^{\mu}(\zeta),w(\zeta)\right]
     =
    \mp\frac{m}{2} \int d\zeta  \left\{
      \frac{\eta_{\mu\nu}\dot{x}^{\mu}\dot{x}^{\nu}}{\dot{w}w}
       \pm m_{\hat{\mu}}\dot{x}^{\hat{\mu}}\right\}\,.
\end{equation}

\noindent
This action is invariant under all the symmetries with the help
of the 1-form $m_{\hat{\mu}}$.

\subsection{$p$-brane Limits}
\label{sec-pbraneADS}

Horospheric coordinates make it easy to take in a very natural way
$p$-brane-type limits of AdS using some of the Cartesian coordinates of the
holographic screen. We will just give a very concise description of this
example: we take the metric Eq.~(\ref{eq:ADShorospheric}) and rescale the
first $(q+1)$ coordinates (we take $q<p$)

\begin{equation}
  x^{A}
  \equiv
  \rho y^{A}\,,
  \hspace{1cm}
  A=0,\cdots,q\,,
\end{equation}

\noindent
obtaining

\begin{equation}
  \label{eq:ADShorospheric3}
    d\hat{s}^{2}_{p+2}
    =
      \hat{h}_{\hat{\mu}\hat{\nu}}d\hat{x}^{\hat{\mu}}d\hat{x}^{\hat{\nu}}
  +\rho^{2}\hat{k}_{\hat{\mu}\hat{\nu}}d\hat{x}^{\hat{\mu}}d\hat{x}^{\hat{\nu}}\,,
\end{equation}

\noindent
where the singular metrics are now

\begin{subequations}
  \begin{align}
    \hat{h}_{\hat{\mu}\hat{\nu}}d\hat{x}^{\hat{\mu}}d\hat{x}^{\hat{\nu}}
    & =
    -\left(\frac{R}{z}\right)^{2}
      \left[\delta_{mn}dx^{m}dx^{n}+dz^{2}\right]\,,
          \hspace{1cm}
    m,n,\ldots
    =
    1,\cdots, p-q\,,
    \\
    & \nonumber \\
    \hat{k}_{\hat{\mu}\hat{\nu}}d\hat{x}^{\hat{\mu}}d\hat{x}^{\hat{\nu}}
    & =
    \left(\frac{R}{z}\right)^{2}
    \left[\eta_{AB}dx^{A}dy^{B}\right]\,.
  \end{align}
\end{subequations}

After  rescaling and relabeling them, the Killing vectors
Eq.~(\ref{eq:ADSKillingvectorshorospheric}) become

\begin{subequations}
  \label{eq:ADSKillingvectorshorospheric4}
  \begin{align}
    \pi_{AB}
    & =
      2\eta_{AB}{}^{C}{}_{D} y^{D}\partial_{C}\,,
    \\
    & \nonumber \\
    \pi_{mn}
    & =
      x^{m}\partial_{n}-x^{n}\partial_{m}\,,
    \\
    & \nonumber \\
    \pi_{+m}
    & =
      \frac{R}{2}\partial_{m}\,,
    \\
    & \nonumber \\
    \pi_{-m}
    & =
      \frac{1}{2R} \left\{
      \left(x\cdot x-z^{2}\right)\partial_{m}
      +2x^{m}(y^{C}\partial_{C}+x^{m}\partial_{m}+z\partial_{z})
      \right\}\,,
    \\
    & \nonumber \\
    \pi
    & =
    \tfrac{1}{2}(y^{C}\partial_{C}+x^{m}\partial_{m}+z\partial_{z})\,,
    \\
    & \nonumber \\
    \varpi_{mA}
    & =
     \frac{1}{\rho}x^{m}\partial_{A} +\rho\eta_{AC}y^{C}\partial_{m}\,,
    \\
    & \nonumber \\
    \varpi_{+A}
    & =
      \frac{1}{\rho}\frac{R}{2}\partial_{A}\,,
    \\
    & \nonumber \\
    \varpi_{-A}
    & =
      \frac{1}{\rho} \frac{1}{2R}\left(x\cdot x-z^{2}\right)\partial_{A}
      -\rho \frac{1}{R}\eta_{AB}y^{B}
      (y^{C}\partial_{C}+x^{m}\partial_{m}+z\partial_{z})\,.
  \end{align}
\end{subequations}

The non-vanishing Lie brackets of these generators are given by

\begin{subequations}
  \label{eq:ADSalgebrahorosphericpbrane}
  \begin{align}
  [\pi_{AB},\pi_{CD}]
  & =
  \eta_{AC}\pi_{BD}
  +\eta_{BD}\pi_{AC}
  -\eta_{AD}\pi_{BC}
  -\eta_{BC}\pi_{AD}\,,
    \\
    & \nonumber \\
  [\pi_{mn},\pi_{pq}]
  & =
  -\delta_{mp}\pi_{nq}
  -\delta_{nq}\pi_{mp}
  +\delta_{mq}\pi_{np}
  +\delta_{np}\pi_{mq}\,,
    \\
    & \nonumber \\
  [\pi_{mn},\pi_{\pm p}]
  & =
    -2\eta_{p[m|}\pi_{\pm|n]}\,,
    \\
    & \nonumber \\
  [\pi_{+ m},\pi_{-n}]
  & =
  \tfrac{1}{2}\pi_{mn}
  +\delta_{mn}\pi\,,
    \\
    & \nonumber \\
  [\pi,\pi_{\pm m}]
  & =
  \pm\tfrac{1}{2}\pi_{\pm m}\,,
    \\
    & \nonumber \\
  [\pi_{AB},\varpi_{ mC}]
  & =
    2\eta_{C[A|}\varpi_{m|B]}\,,
    \\
    & \nonumber \\
  [\pi_{mn},\varpi_{ pA}]
  & =
    -2\delta_{p[m}\varpi_{n]A}\,,
    \\
    & \nonumber \\
  [\pi_{\pm m},\varpi_{ nA}]
  & =
    \delta_{mn}\varpi_{\pm A}\,,
    \\
    & \nonumber \\
  [\pi_{AB},\varpi_{\pm C}]
  & =
    2\eta_{C[A|}\varpi_{\pm|B]}\,,
    \\
    & \nonumber \\
  [\pi,\varpi_{\pm A}]
  & =
  \pm\tfrac{1}{2}\varpi_{\pm A}\,,
    \\
    & \nonumber \\
  [\varpi_{mA} ,\varpi_{ nB}]
  & =
    -\delta_{mn}\pi_{AB}-\eta_{AB}\pi_{mn}\,,
    \\
    & \nonumber \\
  [\varpi_{mA} ,\varpi_{ \pm B}]
  & =
    -\eta_{AB}\pi_{\pm m}\,,
    \\
    & \nonumber \\
  [\varpi_{+ A},\varpi_{- B}]
  & =
  \tfrac{1}{2}\pi_{AB}
  -\eta_{AB}\pi\,.
  \end{align}
\end{subequations}

\noindent
They are of the general form eq.~(\ref{eq:Liealgebrapattern}).

The $\varpi$ generators can be further decomposed as
$\rho^{-1}\varpi^{(-1)} +\rho\varpi^{(1)}$ with

\begin{subequations}
  \begin{align}
        \varpi^{(-1)}_{mA}
    & =
     x^{m}\partial_{A}\,,
    \\
    & \nonumber \\
        \varpi^{(1)}_{mA}
    & =
     \eta_{AC}y^{C}\partial_{m}\,,
    \\
    & \nonumber \\
    \varpi^{(-1)}_{+A}
    & =
      \frac{R}{2}\partial_{A}\,,
    \\
    & \nonumber \\
    \varpi^{(-1)}_{-A}
    & =
     \frac{1}{2R}\left(x\cdot x-z^{2}\right)\partial_{A}\,,
    \\
    & \nonumber \\
    \varpi^{(1)}_{-A}
    & =
      -\frac{1}{R}\eta_{AB}y^{B}
      (y^{C}\partial_{C}+x^{m}\partial_{m}+z\partial_{z})\,.
  \end{align}
\end{subequations}

The generators $\varpi^{(-1)}_{+A}$ will survive both the
$\rho\rightarrow \infty$ and $\rho\rightarrow 0$ limits.  The Lie algebras
obtained in these limits can be found in Table~\ref{yetanothertable}, where it
is explicitly shown how they are related by the general $1\leftrightarrow -1$
duality. This duality does not act on the $\varpi^{(-1)}_{+A}$ generators,
which are common to both algebras. The Lie brackets involving the $\pi$
generators are common to both algebras.

\begin{table}[!ht]
  \centering
  \begin{tabular}{|c|c|}
        \hline
  $\rho\rightarrow \infty$ &  $\rho\rightarrow 0$ \\
    \hline
    & \\
  $[\pi_{AB},\varpi^{(1)}_{ mC}]
  =
    2\eta_{C[A|}\varpi^{(1)}_{m|B]}\,,$
                           &
  $[\pi_{AB},\varpi^{(-1)}_{ mC}]
  =
    2\eta_{C[A|}\varpi^{(-1)}_{m|B]}\,,$
    \\
      & \\
    \hline
    & \\
  $[\pi_{mn},\varpi^{(1)}_{ pA}]
  =
    2\eta_{p[m}\varpi^{(1)}_{n]A}\,,$
                           &
      $[\pi_{mn},\varpi^{(-1)}_{ pA}]
  =
    2\eta_{p[m}\varpi^{(-1)}_{n]A}\,,$
\\
    & \\
    \hline
    & \\
 $[\pi_{- m},\varpi^{(1)}_{ nA}]
  =
    -\eta_{mn}\varpi^{(1)}_{- A}\,,$
                           &
     $[\pi_{- m},\varpi^{(-1)}_{ nA}]
  =
    -\eta_{mn}\varpi^{(-1)}_{- A}\,,$
\\
     \hline
    & \\
   & \\
  $[\pi_{AB},\varpi^{(-1)}_{+C}]
  =
    2\eta_{C[A|}\varpi^{(-1)}_{+|B]}\,,$
                           &
      $[\pi_{AB},\varpi^{(-1)}_{+C}]
  =
    2\eta_{C[A|}\varpi^{(-1)}_{+|B]}\,,$
\\
    & \\
     \hline
    & \\
 $[\pi_{AB},\varpi^{(1)}_{-C}]
  =
    2\eta_{C[A|}\varpi^{(1)}_{-|B]}\,,$
                           &
     $[\pi_{AB},\varpi^{(-1)}_{-C}]
  =
    2\eta_{C[A|}\varpi^{(-1)}_{-|B]}\,,$
\\
    & \\
     \hline
    & \\
 $[\pi,\varpi^{(-1)}_{+ A}]
  =
  \tfrac{1}{2}\varpi^{(-1)}_{+ A}\,,$
                           &
     $[\pi,\varpi^{(-1)}_{+ A}]
  =
  \tfrac{1}{2}\varpi^{(-1)}_{+ A}\,,$
\\
    & \\
     \hline
    & \\
 $[\pi,\varpi^{(1)}_{- A}]
  =
  -\tfrac{1}{2}\varpi^{(1)}_{- A}\,,$
                           &
     $[\pi,\varpi^{(-1)}_{- A}]
  =
  -\tfrac{1}{2}\varpi^{(-1)}_{- A}\,,$
\\
    & \\
    \hline
    & \\
  $[\varpi^{(1)}_{mA} ,\varpi^{(-1)}_{+ B}]
  =
    -\eta_{AB}\pi_{+ m}\,,$
                           &
      $[\varpi^{(-1)}_{mA} ,\varpi^{(-1)}_{+ B}]
  =
0\,,$
\\
    & \\ 
    \hline
    & \\
  $[\varpi^{(-1)}_{+ A},\varpi^{(1)}_{- B}]
  =
  \tfrac{1}{2}\pi_{AB}
  -\eta_{AB}\pi\,,$
                           &
 $[\varpi^{(-1)}_{+ A},\varpi^{(-1)}_{- B}]
  =
0\,,$
                             \\
    & \\
    \hline
  \end{tabular}
  \caption{\small This table gives the non-vanishing Lie brackets that arise
    after taking the $p$-brane $\rho\rightarrow \infty$ and
    $\rho\rightarrow 0$ limits of the AdS algebra.}
  \label{yetanothertable}
\end{table}

In this section we have seen how the general framework that we have developed
allows us to explore systematically new, less conventional singular metrics of
the AdS metric. In the next section we are going to study an even less
conventional limit of a well-known family of solutions of the vacuum Einstein
equations.

\section{Example 3: $pp$-Waves}
\label{sec-pp}

In this last example we are going to consider a 4-dimensional $pp$-wave metric
of the form

\begin{equation}\label{ppwave}
  ds^{2}
  =
  2du(dv+\tfrac{1}{2}Hdu) -dx^{m}dx^{m}\,,
  \,\,\,\,
  m=1,2,
  \hspace{1cm}
  H
  =
  H_{mn}(u)x^{m}x^{n}\,.
\end{equation}

\noindent
(We will also refer to the wavefront coordinates as $x^{1},x^{2}$ as $x$ and
$y$ respectively.)  These metrics admit a null covariantly constant Killing
vector $\partial_{v}$ and solve the vacuum Einstein equations for any
traceless $u$-dependent $2\times 2$ matrix $H_{mn}(u)$. The most general
matrix of this kind can be written in the form

\begin{equation}
  H
  =
  a(u) \sigma^{1}+b(u)\sigma^{3}
  =
  \left(
    \begin{array}{lr}
      b(u)  & a(u) \\
      & \\
      a(u)  & -b(u) \\
     \end{array}
  \right)\,,
\end{equation}

\noindent
and, therefore, are determined by just two functions of $u$: $a(u)$ and
$b(u)$.

Generically, all the metrics Eq.~\eqref{ppwave} have the following 5 Killing
vectors $\{X_{A}\}$ with $A=(a,5)$ and $a=1,\cdots,4$:

\begin{subequations}
  \begin{align}
    X_{a}
    & =
  (f'_{a} x + g'_{a} y )\partial_{v} + f_{a} \partial_{x} + g_{a}\partial_{y}\, ,
    \\
    & \nonumber \\
    X_{5}
    & =
            \partial_{v} \, ,
  \end{align}
\end{subequations}

\noindent
where the prime denotes derivative with respect to $u$ and and the subindex
$a=1,\cdots,4$ labels the 4 solutions of $f$ and $g$ of the following two
differential equations

\begin{equation}
    \label{set}
\begin{aligned}
  f''
  & =
  -b(u) f - a(u) g \,,
  \\
  & \\
  g''
  & =
  b(u) g - a(u) f\,.
\end{aligned}
\end{equation}

\noindent
Furthermore, the commutators of these Killing vectors are given by

\begin{subequations}
  \begin{align}
    [X_{5} , X_{a}]
    & =
      0\,,
    \\
    & \nonumber \\
    [X_{a}, X_{c}]
    & =
      (f'_{c}f_{a} - f'_{a}f_{c} + g'_{c}g_{a} - g'_{a} g_{c}) X_{5}\,.
  \end{align}
\end{subequations}

In the case that $a(u)= a$ and $b(u)= b$ with $a,b$ constant, we can solve
Eqs.~(\ref{set}) obtaining

\begin{subequations}
              \label{fg_constant}
  \begin{align}
    f_{1}
    & =
      a\frac{\cos\left[ \left(a^{2} + b^{2} \right)^{1/4} u\right] -\cosh
      \left[\left( a^{2} + b^{2} \right)^{1/4} u\right] }{2 \sqrt{a^{2} +
      b^{2}}}\,,
    \\
    & \nonumber \\
    g_{1}
    & =
      \frac{\left(-b + \sqrt{a^{2} + b^{2}}\right) \cos\left[\left(a^{2} +
      b^{2} \right)^{1/4} u \right]+ \left(b + \sqrt{a^{2} +
      b^{2}}\right)\cosh\left[\left(a^{2} + b^{2} \right)^{1/4} u \right]}{2
      \sqrt{a^{2} + b^{2}}}\,,
    \\
    & \nonumber \\
    f_{2}
    & =
      a\frac{\sin\left[ \left(a^{2} + b^{2} \right)^{1/4} u\right] -\sinh
      \left[\left( a^{2} + b^{2} \right)^{1/4} u\right] }{2 \left(a^{2} +
      b^{2}\right)^{3/4}}\,,
    \\
    & \nonumber\\
    g_{2}
    & =
      \frac{\left(-b + \sqrt{a^{2} + b^{2}}\right) \sin\left[\left(a^{2} +
      b^{2} \right)^{1/4} u \right]+ \left(b + \sqrt{a^{2} +
      b^{2}}\right)\sinh\left[\left(a^{2} + b^{2} \right)^{1/4} u \right]}{2
      \left(a^{2} + b^{2}\right)^{3/4}}\,,
    \\
    & \nonumber \\
    f_{3}
    & =
      \frac{\left(b + \sqrt{a^{2} + b^{2}}\right) \cos\left[\left(a^{2} +
      b^{2} \right)^{1/4} u \right]+ \left(-b + \sqrt{a^{2} +
      b^{2}}\right)\cosh\left[\left(a^{2} + b^{2} \right)^{1/4} u \right]}{2
      \sqrt{a^{2} + b^{2}}}\,,
      \\
      & \nonumber \\
    g_{3}
    & =
      a\frac{\cos\left[ \left(a^{2} + b^{2} \right)^{1/4} u\right] -\cosh
      \left[\left( a^{2} + b^{2} \right)^{1/4} u\right] }{2 \sqrt{a^{2} +
      b^{2}}}\,,
      \\
      & \nonumber \\
    f_{4}
    & =
      \frac{\left(b + \sqrt{a^{2} + b^{2}}\right) \sin\left[\left(a^{2} +
      b^{2} \right)^{1/4} u \right]+ \left(-b + \sqrt{a^{2} +
      b^{2}}\right)\sinh\left[\left(a^{2} + b^{2} \right)^{1/4} u \right]}{2
      \left(a^{2} + b^{2}\right)^{3/4}}\,,
      \\
      & \nonumber \\
    g_{4}
    & =
      a\frac{\sin\left[ \left(a^{2} + b^{2} \right)^{1/4} u\right] -\sinh \left[\left( a^{2} + b^{2} \right)^{1/4} u\right] }{ 2 \left(a^{2} + b^{2}\right)^{3/4}}\,,
  \end{align}
\end{subequations}

\noindent
and there is an additional Killing vector

\begin{equation}
X_{6} = \partial_{u}\,.
\end{equation}

The non-vanishing commutators of the these 6  Killing vectors are given by

\begin{subequations}
  \begin{align}
    [X_{1}, X_{2}] & = [X_{3},X_{4}] = X_{5}  \,,
    \\
    & \nonumber \\
    [X_{6}, X_{1}] & = b X_{2} - a X_{4} \,,
    \\
    & \nonumber \\
    [X_{6}, X_{2}] & = X_{1}\,,
    \\
    & \nonumber \\
    [X_{6},X_{3}] & = -a X_{2} - b X_{4}\,,
    \\
    & \nonumber \\
    [X_{6}, X_{4}] & =  X_{3}\,,
  \end{align}
\end{subequations}

\noindent
and correspond to the so-called twisted Heisenberg algebra
\cite{Blau:2002js}\footnote{The correspondence between the generators $X_{A}$
  and the generators $X^{(k)}, X^{*(l)}$ in Ref.~\cite{Blau:2002js} is
  $(X_{3},X_{1}) \leftrightarrow (X^{(1)},X^{(2)}) $,
  $(X_{4},X_{2}) \leftrightarrow (X^{*(1)},X^{*(2)}) $,
  $X_{5}\leftrightarrow Z $ and $X_{6}\leftrightarrow X $.}

\begin{subequations}
  \begin{align}
    [X^{(k)}, X^{*(l)}]
    & =
      \delta_{kl} Z  \,,
    \\
    & \nonumber \\
    [X, X^{(k)}]
    & =
      -H_{kl} X^{*(l)}\,,
    \\
    & \nonumber \\
    [X, X^{*(k)}]
    & =
      X^{(k)}\,.
  \end{align}
\end{subequations}

If we rescale the coordinates of the wavefront, $x^{m}$  as

\begin{equation}
x^{m} \rightarrow \rho x^{m} \,,
\end{equation}

\noindent
the metric and its Killing vectors take the form

\begin{subequations}
  \begin{align}
    ds^{2}
    & =
      2 du dv + \rho^{2} (H du^{2} - dx^{m} dx^{m})\,,
    \\
    & \nonumber \\
    X_{a}
    & =
      \rho (f'_{a} x+ g'_{a}y) \partial_{v} + \frac{1}{\rho}
      (f_{a}\partial_{x} + g_{a} \partial_{y})\,,
    \\
    & \nonumber \\
    X_{5}
    & =
      \partial_{v}\,,
    \\
    & \nonumber \\
    X_{6}
    & =
      \partial_{u}\,,
  \end{align}
\end{subequations}

\noindent
fitting in the general pattern studied in Section~\ref{sec-framework} with the
two singular metrics given by

\begin{subequations}
  \begin{align}
    h_{\mu\nu}dx^{\mu}dx^{\nu}
    & =
      2 du dv\,,
    \\
    & \nonumber \\
    k_{\mu\nu}dx^{\mu}dx^{\nu}
    & =
      H du^{2} - dx^{m} dx^{m}\,.
  \end{align}
\end{subequations}

\noindent
Observe that in this case, the ranks of these two metrics add up to 5, not
just 4. While this is unusual, it is certainly not inconsistent in this
framework, although it might be in a more general one. Another unusual feature
of this particular split is that the two singular metrics are
Lorentzian,\footnote{Lorentzian in a restricted 2- and 3-dimensional sense,
  but not in the original 4-dimensional sense.} with signatures $(+-)$ and
$+--$ respectively. The $u$ coordinate is a lightcone coordinate of
$h_{\mu\nu}$ but it is a time coordinate of $k_{\mu\nu}$. If the functions
$a(u)$ and $b(u)$ are not constant, the latter is a dynamical, time-dependent
metric.

Following the general procedure, we define

\begin{subequations}
  \begin{align}
  X_{a}
  & \equiv
  \varpi_{a}
  =
  \rho^{-1}  \varpi^{(-1)}_{a}
    +\rho  \varpi^{(1)}_{a}\,,
    \\
    & \nonumber \\
    X_{5,6}
    & \equiv
      \pi_{5,6}\,,
  \end{align}
\end{subequations}

\noindent
with

\begin{subequations}
  \begin{align}
    \varpi_{a}^{(-1)}
    & =
      f_{a}\partial_{x} + g_{a} \partial_{y} \,,
    \\
    & \nonumber \\
    \varpi_{a}^{(1)}
    & =
      (f'_{a} x + g'_{a}y)\partial_{v} \,.
  \end{align}
\end{subequations}

We can apply our general results straightforwardly to this case finding that,
in the $\rho\rightarrow 0,\infty$ limits, the non-vanishing commutators of the
symmetry algebra of the point-particle action are given by

\begin{subequations}
  \begin{align}
    [\pi_{6}, \varpi_{1}^{(\pm 1)}]
    & =
      b \varpi_{2}^{(\pm 1)} - a \varpi_{4}^{(\pm 1)} \,,
    \\
    & \nonumber \\
    [\pi_{6}, \varpi_{2}^{(\pm 1)}]
    & =
      \varpi_{1}^{(\pm 1)}\,,
    \\
    & \nonumber \\
    [\pi_{6},\varpi_{3}^{(\pm 1)}]
    & =
      -a \varpi_{2}^{(\pm 1)} - b \varpi_{4}^{(\pm 1)}\,,
    \\
    & \nonumber \\
    [\pi_{6}, \varpi_{4}^{(\pm 1)}]
    & =
      \varpi_{3}^{(\pm 1)}\,.
  \end{align}
\end{subequations}

Yet again, the dimension of the symmetry algebra remains unchanged in the two
singular limits, even if in this particular case we have singular metrics
whose ranks add up to 5. The physical interpretation of the actions obtained
after taking the limits according to the different procedures we have
described in Section~\ref{sec-framework} is, generically, that of actions of
particles moving in backgrounds described by one of the two singular metrics,
although both may occur in the action. A detailed, case by case, study of the
equations of motion, solving constraints and eliminating redundant variables
may shed light and help us in getting a better understanding of the physics
involved in these limits.

\section{Discussion}
\label{sec-discussion}

In this work we have constructed a general framework which allows one to
study the fate of

\begin{enumerate}
\item the Killing vectors and the corresponding isometry algebra of metrics
  which depend on an arbitrary parameter $\rho$ as in
  Eq.~(\ref{eq:assumptionmetric})
\item the symmetries and the corresponding symmetry algebra of relativistic
  particles and branes moving in the background of metrics which depend on an
  arbitrary parameter $\rho$ as in Eq.~(\ref{eq:assumptionmetric})
\end{enumerate}

\noindent
in the $\rho\rightarrow 0,\infty$ limits in which the metric is singular or
ill-defined. 

In particular, this general framework allows us to determine the limiting
symmetry algebras in a simple and systematic way, showing that the dimension
of the symmetry algebra is not changed by taking the singular limits and
leading to new relations or dualities between the algebras obtained after
taking the $\rho\rightarrow 0$ limit and the algebras obtained after taking the
$\rho\rightarrow \infty$ limit.  This included two different dualities
between $p$-brane Galilei and Carroll algebras.

At the level of the Lie algebra of symmetries our framework (excluding the
option of adding a WZ term to the sigma model) uses techniques that resemble
the ones that are used in a Lie algebra expansion providing a different
perspective in which it is a background solution and not the structure group
that plays a central role.  This relation with Lie algebra expansions
suggests several extensions of our framework including supersymmetric ones
\cite{Romano:2019ulw} and extending the limit to a general expansion order by
order in the contraction parameter.\,\footnote{For a recent example of a
  Carroll expansion going beyond the limit, see \cite{Hansen:2021fxi}.}

An attractive feature of our general framework is that it can be applied to a
wide range of situations. The parameter that we are using in taking the limit
can be anything, a velocity, a time parameter, a radial parameter etc., as
long as it is in agreement with the properties of the sigma model and solution
we are taking the limit of. In particular, the ``holographic'' limits that we
have been studying may be useful in the context of studying aspects of the
AdS/CFT correspondence. Other limits, using the velocity of light, have been
used to study holography in relation to non-relativistic theories of gravity
in the bulk
\cite{Gomis:2005pg,Hansen:2020pqs,Fontanella:2021btt,Fontanella:2022fjd,deBoer:2023fnj}. Furthermore,
as we have seen in the 4-dimensional $pp$-wave example in which the ranks of
the singular metrics add up to $5$, this setup includes limits which cannot be
handled by other current methods in which, however, the dimensoin of the
symmetry algebra remains invariant in the limit. It would be interesting to
see whether for those cases a Cartan-like formalism with ``inverse'' metrics,
connection and curvature can be defined in this kind of cases.

A prominent question about our framework in need of an answer is the
following: if we start with a metric of the form
Eq.~(\ref{eq:assumptionmetric}) which is a classical solution of General
Relativity, \textit{are the singular metrics obtained in the limits solutions
  of some gravity theory that may also be obtained by taking related limits of
  the Lagrangian or equations of motion of General Relativity?}

Another important question concerning the use of the Hubbard-Stratonovich
transformation when taking the limits of particle and $p$-brane actions is the
following: {\sl can we give a spacetime geometric meaning to the Lagrange
  multipliers that occur in the worldvolume actions after taking some of
  our limits?} One suggestion is that these extra fields naturally arise by
using a form of the sigma model where the equations of motion occur in a
first-order or Hamiltonian formulation.

We hope to come back to these questions in a future publication.

\section*{Acknowledgments}

TO~would like to thank D.~Pere\~niguez and L.~Romano for interesting
conversations and the Van Swinderen Institute for Theoretical Physics at the
University of Groningen for its hospitality and financial support. This work
was started during a visit of E.B.~to the Institute of Particle Physics and
Gravity in Madrid (IFT) where he was supported by the Severo Ochoa Associates
program.  E.B.~wishes to thank the institute for its kind hospitality and for
providing a stimulating atmosphere.  This work has been supported in part by
the MCI, AEI, FEDER (UE) grants PID2021-125700NB-C21 (``Gravity, Supergravity
and Superstrings'' (GRASS)) and IFT Centro de Excelencia Severo Ochoa
CEX2020-001007-S.  TO wishes to thank M.M.~Fern\'andez for her permanent
support.

\appendix

\section{Anti-De Sitter Spacetime}
\label{sec-ADS}

In this appendix we are going to review the definition of the horospheric
coordinates of AdS that we use in Section~\ref{sec-holographiclimit} and we
are also going to derive the explicit form of the Killing vectors in these
coordinates using the isometric embedding of AdS space as a hyperboloid in a
1-dimension higher spacetime with signature $(++,-\cdots,-)$.

Following Gibbons \cite{Gibbons:2011sg}, we are interested in Anti-De Sitter
(AdS) spacetime in $p+2$ dimensions (AdS$_{p+2}$), which typically arises in
the near-horizon limit of extremal $p$-brane solutions. This spacetime and its
isometries can be better understood through its isometric embedding in a
$(p+3)$-dimensional space with the metric

\begin{equation}
  d\hat{\hat{s}}^{2}_{p+3}
  =
  (dX^{-1})^{2}+ (dX^{0})^{2} - d\vec{X}_{p}^{2} - (dX^{p+1})^{2}\,,
  \hspace{1cm}
  \vec{X}_{p}= (X^{1},\cdots,X^{p})\,,
\end{equation}

\noindent
of  signature $(2,p+1)$ as the hyperboloid given by the equation

\begin{equation}
  (X^{-1})^{2}+(X^{0})^{2} - \vec{X}_{p}^{2} - (X^{p+1})^{2}
  =
  R^{2}\,,
\end{equation}

\noindent
where $R$ if the AdS radius.

It is convenient to introduce the $(p+3)$-dimensional, constant,
pseudo-Riemannian metric of signature $(2,p+1)$
$\hat{\hat{\eta}}=$diag$(+,+,-,\cdots,-)$, and the $(p+3)$ coordinates
$(\hat{\hat{X}}^{\hat{\hat{\mu}}})= (X^{-1},X^{0},\vec{X}_{p},X^{p+1})$ in
terms of which the above equations take the simple form

\begin{subequations}
  \begin{align}
  \label{eq:p+3metric}
    d\hat{\hat{s}}^{2}_{p+3}
    & =
    \hat{\hat{\eta}}_{\hat{\hat{\mu}}\hat{\hat{\nu}}}
      d\hat{\hat{X}}^{\hat{\hat{\mu}}}d\hat{\hat{X}}^{\hat{\hat{\nu}}}\,,
    \\
    & \nonumber \\
    \label{eq:hyperboloid}
    R^{2}
    & =
      \hat{\hat{\eta}}_{\hat{\hat{\mu}}\hat{\hat{\nu}}}
      \hat{\hat{X}}^{\hat{\hat{\mu}}}\hat{\hat{X}}^{\hat{\hat{\nu}}}\,,
  \end{align}
\end{subequations}

\noindent
which make evident the $SO(2,p+1)$ invariance of AdS$_{p+2}$. The
finite $SO(2,p+1)$ transformations act on the coordinates as

\begin{equation}
  \hat{\hat{X}}^{\prime\, \hat{\hat{\mu}}}
  \hat{\hat{\Lambda}}^{\hat{\hat{\mu}}}{}_{\hat{\hat{\nu}}}
  \hat{\hat{X}}^{\hat{\hat{\nu}}}\,,
\end{equation}

\noindent
where

\begin{equation}
  \label{eq:SO2p+1condition}
   \hat{\hat{\eta}}_{\hat{\hat{\alpha}}\hat{\hat{\beta}}}
  \hat{\hat{\Lambda}}^{\hat{\hat{\alpha}}}{}_{\hat{\hat{\mu}}}
  \hat{\hat{\Lambda}}^{\hat{\hat{\beta}}}{}_{\hat{\hat{\nu}}}
  =
   \hat{\hat{\eta}}_{\hat{\hat{\mu}}\hat{\hat{\nu}}}\,.
\end{equation}

\noindent
Infinitesimally,

\begin{equation}
  \hat{\hat{X}}^{\prime\, \hat{\hat{\mu}}}
  \left(
    \delta^{\hat{\hat{\mu}}}{}_{\hat{\hat{\nu}}}
  +\hat{\hat{\sigma}}^{\hat{\hat{\mu}}}{}_{\hat{\hat{\nu}}} \right)
  \hat{\hat{X}}^{\hat{\hat{\nu}}}\,,
\end{equation}

\noindent
and the condition Eq.~(\ref{eq:SO2p+1condition}) becomes

\begin{equation}
  \hat{\hat{\sigma}}_{(\hat{\hat{\mu}}\hat{\hat{\nu}})}
  =
  0\,,
  \,\,\,\,\,
  \text{with}
  \,\,\,\,\,
  \hat{\hat{\sigma}}_{\hat{\hat{\mu}}\hat{\hat{\nu}}}
  =
  \hat{\hat{\eta}}_{\hat{\hat{\mu}}\hat{\hat{\rho}}}
  \hat{\hat{\sigma}}^{\hat{\hat{\rho}}}{}_{\hat{\hat{\nu}}}\,.
\end{equation}

It is natural to write the infinitesimal transformations as follows:

\begin{equation}
\delta_{\hat{\hat{\sigma}}} \hat{\hat{X}}^{\hat{\hat{\mu}}}
=
\tfrac{1}{2}\hat{\hat{\sigma}}^{\hat{\hat{\alpha}}\hat{\hat{\beta}}}
\hat{\hat{k}}_{{\hat{\hat{\alpha}}\hat{\hat{\beta}}}}{}^{\hat{\hat{\mu}}}\,,
\end{equation}

\noindent
where

\begin{equation}
  \hat{\hat{k}}_{{\hat{\hat{\alpha}}\hat{\hat{\beta}}}}{}^{\hat{\hat{\mu}}}
  =
  2
  \hat{\hat{\eta}}_{\hat{\hat{\alpha}}\hat{\hat{\beta}}}{}^{\hat{\hat{\mu}}}{}_{\hat{\hat{\nu}}}
  \hat{\hat{X}}^{\hat{\hat{\nu}}}\,,
  \,\,\,\,\,
  \text{with}
  \,\,\,\,\,
  \hat{\hat{\eta}}_{\hat{\hat{\alpha}}\hat{\hat{\beta}}}{}^{\hat{\hat{\mu}}\hat{\hat{\nu}}}
  =
  \tfrac{1}{2}
  \left(\hat{\hat{\eta}}_{\hat{\hat{\alpha}}}{}^{\hat{\hat{\mu}}}
  \hat{\hat{\eta}}_{\hat{\hat{\beta}}}{}^{\hat{\hat{\nu}}}
    -
    \hat{\hat{\eta}}_{\hat{\hat{\alpha}}}{}^{\hat{\hat{\nu}}}
  \hat{\hat{\eta}}_{\hat{\hat{\beta}}}{}^{\hat{\hat{\mu}}}
\right)\,,
\end{equation}

\noindent
are the infinitesimal generators of those transformations.
$\hat{\hat{k}}_{\hat{\hat{\alpha}}\hat{\hat{\beta}}} =
\hat{\hat{k}}_{\hat{\hat{\alpha}}\hat{\hat{\beta}}}{}^{\hat{\hat{\mu}}}\partial_{\hat{\hat{\mu}}}$
are Killing vectors of the $(p+3)$-dimensional metric
Eq.~(\ref{eq:p+3metric}).\footnote{The $(p+3)$-dimensional metric is also
  invariant under constant translations of the coordinates, but these
  transformations do not leave invariant the equation of the hyperboloid
  Eq.~(\ref{eq:hyperboloid}).}

We can associate a generator
$\hat{\hat{M}}_{\hat{\hat{\mu}}\hat{\hat{\nu}}}=
-\hat{\hat{M}}_{\hat{\hat{\nu}}\hat{\hat{\mu}}}$ of the $so(2,p+1)$ Lie
algebra to (minus) each of the Killing vectors
$\hat{\hat{k}}_{\hat{\hat{\mu}}\hat{\hat{\nu}}}$.\footnote{The minus sign is
  due to the relation between the transformations of the coordinates $x^{\mu}$ and the
  transformations of tensor fields $T$ under the infinitesimal
  diffeomorphisms generated by some vector field $\xi$:
  \begin{equation}
    \begin{aligned}
      \delta_{\xi} x^{\mu}
      & =
      \xi^{\mu}\,,
      \\
      \delta_{\xi} T
      & =
      -\pounds_{\xi}  T
      =
       \pounds_{-\xi}  T\,,
    \end{aligned}
  \end{equation}
  where $\pounds_{\xi}$ is the Lie derivative with respect to the vector
  fields $\xi$.  } The structure constants that occur in the commutators of
the generators are minus the structure constants that occur in the Lie
brackets of the vectors

\begin{equation}
  [\hat{\hat{k}}_{\hat{\hat{\mu}}\hat{\hat{\nu}}},\hat{\hat{k}}_{\hat{\hat{\rho}}\hat{\hat{\sigma}}}]
  =
  -\hat{\hat{\eta}}_{\hat{\hat{\mu}}\hat{\hat{\rho}}}\hat{\hat{k}}_{\hat{\hat{\nu}}\hat{\hat{\sigma}}}
  -\hat{\hat{\eta}}_{\hat{\hat{\nu}}\hat{\hat{\sigma}}}\hat{\hat{k}}_{\hat{\hat{\mu}}\hat{\hat{\rho}}}
  +\hat{\hat{\eta}}_{\hat{\hat{\mu}}\hat{\hat{\sigma}}}\hat{\hat{k}}_{\hat{\hat{\nu}}\hat{\hat{\rho}}}
  +\hat{\hat{\eta}}_{\hat{\hat{\nu}}\hat{\hat{\rho}}}\hat{\hat{k}}_{\hat{\hat{\mu}}\hat{\hat{\sigma}}}\,,
\end{equation}

\noindent
and, in this case, one arrives at the following
commutation relations between the generators:

\begin{equation}
  [\hat{\hat{M}}_{\hat{\hat{\mu}}\hat{\hat{\nu}}},\hat{\hat{M}}_{\hat{\hat{\rho}}\hat{\hat{\sigma}}}]
  =
  \hat{\hat{\eta}}_{\hat{\hat{\mu}}\hat{\hat{\rho}}}\hat{\hat{M}}_{\hat{\hat{\nu}}\hat{\hat{\sigma}}}
  +\hat{\hat{\eta}}_{\hat{\hat{\nu}}\hat{\hat{\sigma}}}\hat{\hat{M}}_{\hat{\hat{\mu}}\hat{\hat{\rho}}}
  -\hat{\hat{\eta}}_{\hat{\hat{\mu}}\hat{\hat{\sigma}}}\hat{\hat{M}}_{\hat{\hat{\nu}}\hat{\hat{\rho}}}
  -\hat{\hat{\eta}}_{\hat{\hat{\nu}}\hat{\hat{\rho}}}\hat{\hat{M}}_{\hat{\hat{\mu}}\hat{\hat{\sigma}}}\,.
\end{equation}

\subsection{Horospheric coordinates}
\label{sec-horospheric}

First, we define the ``lightcone'' coordinates

\begin{equation}
\hat{\hat{X}}^{\pm} \equiv \hat{\hat{X}}^{-1}\pm \hat{\hat{X}}^{p+1}\,,
\end{equation}

\noindent
in terms of which the metric and the hyperboloid equation
Eqs.~(\ref{eq:p+3metric}) and (\ref{eq:hyperboloid}) take the form

\begin{subequations}
  \begin{align}
  \label{eq:p+3metric2}
    d\hat{\hat{s}}^{2}_{p+3}
    & =
      d\hat{\hat{X}}^{+}d\hat{\hat{X}}^{-} +\eta_{\mu\nu}dX^{\mu}dX^{\nu}\,,
    \\
    & \nonumber \\
    \label{eq:hyperboloid2}
    R^{2}
    & =
      \hat{\hat{X}}^{+}\hat{\hat{X}}^{-} +\eta_{\mu\nu}X^{\mu}X^{\nu}\,,
  \end{align}
\end{subequations}

\noindent
where we are using the $(p+1)$-dimensional indices $\mu,\nu=0,\cdots,p$ and
Lorentzian metric $(\eta)=$diag$(+,-,\cdots,-)$. We can use the hyperboloid
equation to solve for the coordinate $ \hat{\hat{X}}^{+}$

\begin{equation}
  \hat{\hat{X}}^{+}
  =
  R^{2}-\eta_{\mu\nu}X^{\mu}X^{\nu}/ \hat{\hat{X}}^{-}\,,
\end{equation}

\noindent
and eliminate it from the metric to obtain the $(p+2)$-dimensional metric
induced on the hyperboloid

\begin{equation}
    d\hat{s}^{2}_{p+2}
    =
    -\left(\frac{d\hat{\hat{X}}^{-}}{\hat{\hat{X}}^{-}}\right)^{2}
    \left[R^{2}-\eta_{\mu\nu}X^{\mu}X^{\nu}\right]
      -2\frac{d\hat{\hat{X}}^{-}}{\hat{\hat{X}}^{-}}\eta_{\mu\nu}X^{\mu}dX^{\nu} +\eta_{\mu\nu}dX^{\mu}dX^{\nu}\,.
\end{equation}

The transformation to $(p+2)$-dimensional \textit{horospheric coordinates}
$(\hat{x}^{\hat{\mu}})=(x^{\mu},z)$

\begin{equation}
  \label{eq:horosphericcoordinates}
  \hat{\hat{X}}^{-} \equiv R^{2}/z\,,
  \hspace{1cm}
  X^{\mu} \equiv Rx^{\mu}/z\,,
\end{equation}

\noindent
diagonalizes the metric

\begin{equation}
  \label{eq:ADShorospheric}
    d\hat{s}^{2}_{p+2}
    =
    \left(\frac{R}{z}\right)^{2}
    \left[\eta_{\mu\nu}dx^{\mu}dx^{\nu}-dz^{2}\right]\,.
\end{equation}

By construction, this metric has the same isometry group as the original
$(p+3)$-dimensional one. The Killing vectors have a different form,
though. The simplest way to find them is to take the pullback of the dual
1-forms over the hyperboloid and, then, change the coordinates. The dual
1-forms are given by

\begin{equation}
    \tilde{\hat{\hat{k}}}_{{\hat{\hat{\alpha}}\hat{\hat{\beta}}}}
    = 2
    \hat{\hat{\eta}}_{\hat{\hat{\alpha}}\hat{\hat{\beta}},\hat{\hat{\mu}}\hat{\hat{\nu}}}
    \hat{\hat{X}}^{\hat{\hat{\nu}}} d \hat{\hat{X}}^{\hat{\hat{\mu}}}\,,
\end{equation}

\noindent
and their pullbacks over the hyperboloid are given by

\begin{equation}
  \begin{aligned}
    \tilde{\hat{\hat{k}}}_{{\hat{\hat{\alpha}}\hat{\hat{\beta}}}}
    & = 2
    \hat{\hat{\eta}}_{\hat{\hat{\alpha}}\hat{\hat{\beta}},\mu\nu}
    X^{\nu}dX^{\mu}
    +2\hat{\hat{\eta}}_{\hat{\hat{\alpha}}\hat{\hat{\beta}},\mu -}
    \left(X^{\mu} d \hat{\hat{X}}^{-} -\hat{\hat{X}}^{-} dX^{\mu}\right)
    \\
    & \\
    & \hspace{.5cm}
    +2\hat{\hat{\eta}}_{\hat{\hat{\alpha}}\hat{\hat{\beta}},\mu +}
    \left(X^{\mu} d \hat{\hat{X}}^{+} -\hat{\hat{X}}^{+} dX^{\mu}\right)
    +2\hat{\hat{\eta}}_{\hat{\hat{\alpha}}\hat{\hat{\beta}},+ -}
    \left(\hat{\hat{X}}^{+} d \hat{\hat{X}}^{-} -\hat{\hat{X}}^{-}
      d\hat{\hat{X}}^{+}\right)
    \\
    & \\
    & = 2
    \hat{\hat{\eta}}_{\hat{\hat{\alpha}}\hat{\hat{\beta}},\mu\nu}
    X^{\nu}dX^{\mu}
    +2\hat{\hat{\eta}}_{\hat{\hat{\alpha}}\hat{\hat{\beta}},\mu -}
    \left(X^{\mu} d \hat{\hat{X}}^{-} -\hat{\hat{X}}^{-} dX^{\mu}\right)
    \\
    & \\
    & \hspace{.5cm}
    +2\hat{\hat{\eta}}_{\hat{\hat{\alpha}}\hat{\hat{\beta}},\mu +}
    \left[-\frac{X^{\mu}\left(R^{2}-X\cdot X\right)d\hat{\hat{X}}^{-}}{(\hat{\hat{X}}^{-})^{2}}
      -2\frac{X^{\mu}X\cdot dX}{\hat{\hat{X}}^{-}}
      -\frac{\left(R^{2}-X\cdot X\right)dX^{\mu}}{\hat{\hat{X}}^{-}}\right]
    \\
    & \\
    & \hspace{.5cm}
    +4\hat{\hat{\eta}}_{\hat{\hat{\alpha}}\hat{\hat{\beta}},+ -}
    \left[\frac{\left(R^{2}-X\cdot X\right) d\hat{\hat{X}}^{-}}{\hat{\hat{X}}^{-}}
      +X\cdot dX    \right]\,.
  \end{aligned}
\end{equation}

\noindent
In horospheric coordinates

\begin{equation}
  \begin{aligned}
    \tilde{\hat{\hat{k}}}_{{\hat{\hat{\alpha}}\hat{\hat{\beta}}}}
    & = 2
    \hat{\hat{\eta}}_{\hat{\hat{\alpha}}\hat{\hat{\beta}},\mu\nu}
    \left(\frac{R}{z}\right)^{2}x^{\nu}dx^{\mu}
    \\
    & \\
    & \hspace{.5cm}
    -2\hat{\hat{\eta}}_{\hat{\hat{\alpha}}\hat{\hat{\beta}},\mu -}
    R\left(\frac{R}{z}\right)^{2}
    dx^{\mu}
    \\
    & \\
    & \hspace{.5cm}
    +\frac{2}{R}\hat{\hat{\eta}}_{\hat{\hat{\alpha}}\hat{\hat{\beta}},\mu +}\left(\frac{R}{z}\right)^{2}
    \left[ \left(x\cdot x-z^{2}\right)dx^{\mu}
      -2x^{\mu} x\cdot dx +2zx^{\mu}dz\right]
    \\
    & \\
    & \hspace{.5cm}
    +4\hat{\hat{\eta}}_{\hat{\hat{\alpha}}\hat{\hat{\beta}},+ -}\left(\frac{R}{z}\right)^{2}
    \left(x\cdot dx -zdz   \right)\,.
  \end{aligned}
\end{equation}

Finally, we just have to raise the 1-form index using the inverse of the
metric Eq.~(\ref{eq:ADShorospheric}) to find the vectors. Observe that this
removes the overall $(R/z)^{2}$ factor. The result is

\begin{equation}
  \begin{aligned}
    \hat{\hat{k}}_{{\hat{\hat{\alpha}}\hat{\hat{\beta}}}}
    & = 2
    \hat{\hat{\eta}}_{\hat{\hat{\alpha}}\hat{\hat{\beta}}}{}^{\mu}{}_{\nu} x^{\nu}\partial_{\mu}
    -2R\hat{\hat{\eta}}_{\hat{\hat{\alpha}}\hat{\hat{\beta}}}{}^{\mu}{}_{-}\partial_{\mu}
    \\
    & \\
    & \hspace{.5cm}
    +\frac{2}{R}\hat{\hat{\eta}}_{\hat{\hat{\alpha}}\hat{\hat{\beta}}}{}^{\mu}{}_{+}
     \left(x\cdot x-z^{2}\right)\partial_{\mu}
        -\frac{4}{R}\hat{\hat{\eta}}_{\hat{\hat{\alpha}}\hat{\hat{\beta}},\mu +}
 x^{\mu} x^{\nu}\partial_{\nu}
     -\frac{4}{R}\hat{\hat{\eta}}_{\hat{\hat{\alpha}}\hat{\hat{\beta}},\mu +}
    zx^{\mu}\partial_{z}
   \\
    & \\
    & \hspace{.5cm}
    +4\hat{\hat{\eta}}_{\hat{\hat{\alpha}}\hat{\hat{\beta}},+ -}
    \left(x^{\mu}\partial_{\mu}+z\partial_{z}\right)\,.
  \end{aligned}
\end{equation}

Now, taking into account that the non-vanishing components of
$\hat{\hat{\eta}}$ in the lightcone basis are $\hat{\hat{\eta}}_{+-}=1/2$ and
$\hat{\hat{\eta}}_{\mu\nu}=\eta_{\mu\nu}$, we find the following independent
Killing vectors:

\begin{subequations}
  \label{eq:ADSKillingvectorshorospheric}
  \begin{align}
    \hat{\hat{k}}_{\alpha\beta}
    & =
      2\eta_{\alpha\beta}{}^{\mu}{}_{\nu} x^{\nu}\partial_{\mu}\,,
    \\
    & \nonumber \\
    \hat{\hat{k}}_{+\alpha}
    & =
      \frac{R}{2}\partial_{\alpha}\,,
    \\
    & \nonumber \\
    \hat{\hat{k}}_{-\alpha}
    & =
      \frac{1}{2R} \left\{
      \left(x\cdot x-z^{2}\right)\partial_{\alpha}
      -2\eta_{\alpha\mu}x^{\mu}(x^{\nu}\partial_{\nu}+z\partial_{z})
      \right\}\,,
    \\
    & \nonumber \\
    \hat{\hat{k}}_{-+}
    & =
    \tfrac{1}{2}\left(  x^{\mu}\partial_{\mu}+z\partial_{z}\right)\,.
  \end{align}
\end{subequations}

The non-vanishing commutation relations of the corresponding generators are

\begin{subequations}
  \label{eq:ADSalgebrahorospheric}
  \begin{align}
  [\hat{\hat{M}}_{\mu\nu},\hat{\hat{M}}_{\rho\sigma}]
  & =
  \eta_{\mu\rho}\hat{\hat{M}}_{\nu\sigma}
  +\eta_{\nu\sigma}\hat{\hat{M}}_{\mu\rho}
  -\eta_{\mu\sigma}\hat{\hat{M}}_{\nu\rho}
  -\eta_{\nu\rho}\hat{\hat{M}}_{\mu\sigma}\,,
    \\
    & \nonumber \\
  [\hat{\hat{M}}_{\pm\alpha},\hat{\hat{M}}_{\mu\nu}]
  & =
    -2\eta_{\alpha[\mu|}\hat{\hat{M}}_{\pm|\nu]}\,,
    \\
    & \nonumber \\
  [\hat{\hat{M}}_{+\alpha},\hat{\hat{M}}_{-\beta}]
  & =
  \tfrac{1}{2}\hat{\hat{M}}_{\alpha\beta}
  -\eta_{\alpha\beta}\hat{\hat{M}}_{-+}\,,
    \\
    & \nonumber \\
  [\hat{\hat{M}}_{\pm\alpha},\hat{\hat{M}}_{-+}]
  & =
  \mp\tfrac{1}{2}\hat{\hat{M}}_{\pm\alpha}\,.
  \end{align}
\end{subequations}

Defining\footnote{If $P_{\alpha}$ are rescaled with inverse factors the
  structure constants of the algebra remain invariant.}

\begin{equation}
  \hat{\hat{M}}_{\mu\nu}
  \equiv
  M_{\mu\nu}\,,
\hspace{.5cm}
  \hat{\hat{M}}_{+\alpha} \equiv \frac{R}{2} P_{\alpha}\,,
  \hspace{.5cm}
  \hat{\hat{M}}_{-\alpha} \equiv \frac{1}{2R} B_{\alpha}\,,
  \hspace{.5cm}
  \hat{\hat{M}}_{-+}
  \equiv \tfrac{1}{2} D\,,
\end{equation}

\noindent
the above algebra takes the form

\begin{subequations}
  \label{eq:adsp+2algebraso(1,p)split}
  \begin{align}
  [M_{\mu\nu},M_{\rho\sigma}]
  & =
  \eta_{\mu\rho}M_{\nu\sigma}
  +\eta_{\nu\sigma}M_{\mu\rho}
  -\eta_{\mu\sigma}M_{\nu\rho}
  -\eta_{\nu\rho}M_{\mu\sigma}\,,
    \\
    & \nonumber \\
  [P_{\alpha},M_{\mu\nu}]
  & =
    -2\eta_{\alpha[\mu|}P_{|\nu]}\,,
    \\
    & \nonumber \\
  [B_{\alpha},M_{\mu\nu}]
  & =
    -2\eta_{\alpha[\mu|}B_{|\nu]}\,,
    \\
    & \nonumber \\
  [P_{\alpha},B_{\beta}]
  & =
  2 M_{\alpha\beta}
  -2\eta_{\alpha\beta}D\,,
    \\
    & \nonumber \\
  [P_{\alpha},D]
  & =
  -P_{\alpha}\,,
    \\
    & \nonumber \\
  [B_{\alpha},D]
  & =
  B_{\alpha}\,.
  \end{align}
\end{subequations}


\end{document}